\documentclass{iopjournal}

\usepackage{amsmath}
\usepackage{amssymb}
\usepackage{graphicx}
\usepackage{amsfonts}
\usepackage{color}
\usepackage[sort&compress]{natbib}
\usepackage{chemformula}
\usepackage{ragged2e}
\usepackage{hyperref}

\AtBeginDocument{\justifying}
\usepackage{soul}
\usepackage{physics}

\begin{document}

\articletype{Paper} 
\justifying

\title{Universally Diverging Gr\"uneisen Ratio of Holographic Quantum Criticality}

\author{
Jun-Kun Zhao$^{1,\dagger}$\orcid{0000-0002-5835-3920},
Enze Lv$^{1,2,\dagger}$\orcid{0009-0009-6140-3069}, 
Wei Li$^{1,2,*}$\orcid{0000-0003-0474-9337} and 
Li Li$^{1,2,3,*}$\orcid{0000-0003-3124-5281} }

\affil{$^1$Institute of Theoretical Physics, Chinese Academy of Sciences, Beijing 100190, China}

\affil{$^2$School of Physical Sciences, University of Chinese Academy of Sciences, Beijing 100049, China}

\affil{$^3$School of Fundamental Physics and Mathematical Sciences, Hangzhou Institute for Advanced Study, UCAS, Hangzhou 310024, China}

\affil{$^\dagger$These authors contributed equally to this work.}\\
\affil{$^*$Authors to whom any correspondence should be addressed.}

\email{w.li@itp.ac.cn; liliphy@itp.ac.cn}


\begin{abstract}
Quantum criticality is a hallmark of strongly correlated electron systems, as seen in heavy-fermion materials and high-temperature superconductors. Holographic duality provides a powerful framework to investigate these systems by translating them into weakly coupled classical gravity living in one higher dimension. Here, we harness this approach to study a field-induced quantum critical point with dynamical exponent $z=3$ in Einstein-Maxwell-Chern-Simons theory. Our analysis of its thermodynamic properties reveals a new universality class. Notably, we identify a diverging Gr\"uneisen ratio with universal scaling $\sim T^{-2/3}$, a behavior that closely mirrors recent experiments on the heavy-fermion material~CeRh$_6$Ge$_4$. These findings advance our understanding of metallic quantum criticality and highlight the potential of holographic duality as a tool for studying correlated quantum matters.
\end{abstract}

\section{Introduction}

Quantum phase transitions (QPTs) occur at zero temperature in strongly correlated systems and are driven by the competition between quantum fluctuations and interactions~\cite{Sondhi1997,sachdev2015}. A continuous QPT features a quantum critical point (QCP) that separates different phases of matter. At the QCP, the system exhibits scale invariance and gives rise to nontrivial low-energy excitations that defy conventional quasiparticle descriptions. The influence of a QCP extends across a wide range of temperatures and tuning parameters, leading to universal scaling behavior in a region known as the quantum critical regime (QCR)~\cite{sachdev2000,Coleman2005}. Over recent decades, quantum criticality has emerged as a central theme connecting condensed matter physics, quantum information, and quantum gravity, with significant theoretical and experimental progress~\cite{sachdev2000,Coleman2005,Sebastian2006Purple,Gegenwart2007,coldea2010,Cubrovic:2009ye,Xiang2024GiantME}.

Near the QCP, thermodynamics and response function exhibit universal quantum scaling laws due to the scale invariance of the system~\cite{sachdev2015}. A prominent example is the Gr\"uneisen ratio, $\Gamma_g \equiv {1}/{T} \left({\partial T}/{\partial g}\right)_S$, which characters the adiabatic temperature change under external field $g$~\cite{Gruneisen1912, Wolf2014CoolingTQ, Xiang2024GiantME, Shu2026Nature}. At the QCP, the Gr\"uneisen ratio universally diverges as $\Gamma_g\propto T^{-1/z\nu}$ with $z$ the dynamic critical exponent and $\nu$ the critical exponent of correlation length~\cite{Zhu2003UniversallyDG,Garst2005SignCO}. In experiments, this Gr\"uneisen ratio scaling of $T^{-1/z\nu}$ has been observed in quantum magnets~\cite{sciadvaao3773,xiang2025} and heavy-fermion materials~\cite{Kuchler2003, Kuchler2004, Kuchler_2007, Tokiwa2009, Alexander2013, Gegenwart2016GrneisenPS, Bin2020, Zhan2025CriticalFA}. It thereby serves as a highly sensitive probe for detecting the QCP and determining its critical exponents~\cite{Zhu2003UniversallyDG, Garst2005SignCO, xiang2025, Zhan2025CriticalFA}. 

Recent studies on heavy-fermion material~CeRh$_6$Ge$_4$ reveal a diverging Gr\"uneisen ratio $\Gamma_P\propto T^{-2/3}$ under external pressure $P$~\cite{Bin2020,Zhan2025CriticalFA}, implying a novel QCP with cubic dispersion $z=3$ and critical exponent $\nu=1/2$. This continuous QPT goes beyond the Landau paradigm~\cite{Zhan2025CriticalFA}, and a theoretical understanding of the metallic QCP remains a formidable challenge. The seminal effective field theory framework developed by Hertz and Millis provides an important approach to such metallic quantum criticality~\cite{Hertz:1976zz,Millis:1993}. However, by integrating out the fermions to derive an effective action for the order parameter, this approach neglects the crucial backreaction of order parameter fluctuations on the fermionic degrees of freedom---a procedure that is dangerous~\cite{Lohneysen:2007zz}. Consequently, the specific universality class governing these quantum critical phenomena still requires further scrutiny. Actually, calculating metallic systems with both spin and charge degrees of freedom remains a challenging problem for quantum many-body methods. Quantum Monte Carlo simulations suffer from the sign problem, while tensor network methods face difficulties in increasing the bond dimension.

Here, employing holographic duality, we investigate the metallic QCP and gain new insights into its quantum scaling and universal behavior. Holography, or the Anti-de Sitter/Conformal Field Theory (AdS/CFT) correspondence, establishes a duality between a $(d+1)$-dimensional gravitational theory and a $d$-dimensional quantum many-body system living on its boundary~~\cite{Maldacena:1997re,Witten:1998qj,Gubser:1998bc}. Holographic methods have proven successfully in modeling strongly correlated systems without quasiparticle excitations, capturing not only conventional Fermi-liquid behavior but also exotic non-Fermi-liquid physics observed in strange metals~\cite{Liu:2009dm,Cubrovic:2009ye,Faulkner:2009wj,Horowitz:2012ky}. Notably, the emergent AdS$_2\times \mathbb{R}^{d-1}$ near-horizon geometry in the $(d+1)$ extremal Reissner-N\"ordstorm–AdS black hole is responsible for the non-Fermi-liquid behavior in fermionic correlation functions~\cite{Faulkner:2009wj}, and admits a dimensional reduction to the two-dimensional Jackiw–Teitelboim gravity, revealing a low-energy duality to the Sachdev–Ye–Kitaev model~\cite{Sachdev1992GaplessSG,Maldacena:2016hyu,Maldacena:2016upp}. Furthermore, holographic approaches provide a novel framework for realizing superconductivity or superfluidity~\cite{Hartnoll:2008vx,Adams:2012pj,Cai:2017qdz}, offer theoretical laboratories for exploring far-from-equilibrium dynamics~\cite{Chesler:2008hg,Adams:2013vsa,bhaseen2015,Baggioli:2021tzr}, and reveal the intrinsic link between quantum information and gravity~\cite{Ryu:2006bv,Takayanagi:2017knl,Takayanagi:2025ula}. For comprehensive reviews, see \emph{e.g.}~\cite{Zaanen:2015oix,Ammon:2015wua,Cai:2015cya,Hartnoll:2018xxg,Baggioli:2021xuv}.

\begin{figure} 
\centering
\includegraphics[width=0.8\textwidth]{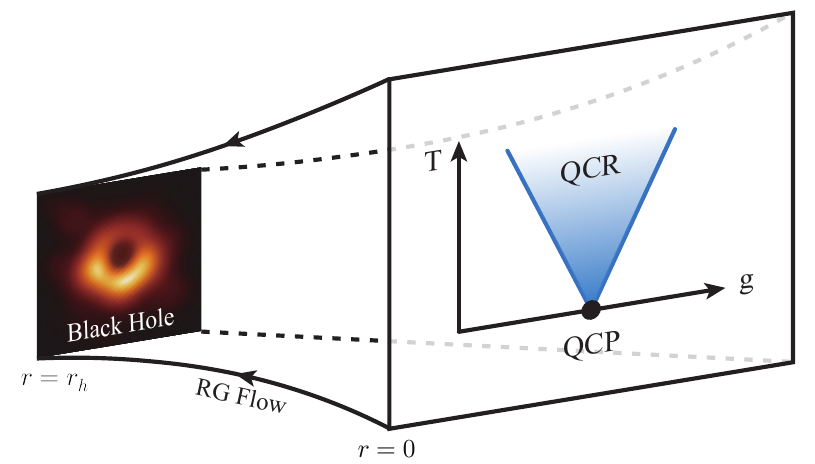} 
\caption{\textbf{Holographic description of quantum criticality.} 
The dual quantum many-body system lives at the boundary ($r=0$) of the bulk gravity system. The holographic radial coordinate $r$ geometries the energy scale or the renormalization group (RG) flow of the boundary system: processes near the boundary ($r\sim0$) describe the short-distance, ultraviolet dynamics of the dual system, while processes in the interior ($r\gg0$) capture its universal low-energy dynamics. Within this holographic framework, operators in the boundary quantum field theory are dual to classical fields in the bulk, where external sources for these operators are encoded in the asymptotic behavior of the corresponding bulk fields. QPTs are realized by tuning the asymptotic boundary conditions of the gravitational theory, which leads to distinct scaling geometries in the deep infrared regions. The finite temperature physics of the boundary QPT, including its QCR, can be obtained by heating up the zero temperature scaling geometries, \emph{i.e.} by considering a black hole located at $r=r_h$.}\label{fig:adscft}
\end{figure}

QPT is holographically described as follows: by continuously tuning a non-thermal parameter, the system evolves from an ultraviolet fixed point toward distinct infrared fixed points—realized as different scaling geometries, as illustrated in Fig.~\ref{fig:adscft}. Notable examples of holographic QPTs include cohesive-fractionalized transition~\cite{Hartnoll:2011pp,Hartnoll:2012ux}, metal-insulator transition~\cite{Donos:2012js}, Weyl semimetal~\cite{Landsteiner:2015pdh,Ammon:2016mwa,Liu:2018spp}, etc. Of particular interest is the five dimensional Einstein-Maxwell-Chern-Simons (EMCS) theory~\cite{Erdmenger:2008rm,Banerjee:2008th,Son:2009tf,Donos:2012wi,Rangamani:2023mok}, which is simple and universal in the sense that it only contains metric and gauge fields yet can give rise to novel magnetic field-induced QPT~\cite{DHoker:2010zpp,DHoker:2010onp,DHoker:2012rlj}. One appealing property of this model is that it hosts a $z=3$ QCP, offering a unique theoretical platform to investigate this notable class of quantum criticality observed in heavy-fermion materials such as~CeRh$_6$Ge$_4$. In this work, we investigate the quantum criticality near this metallic QCP, aiming to offer a new perspective for understanding analogous behavior in heavy-fermion materials. Although quantum criticality has been extensively studied within the holographic framework, to the best of our knowledge, a systematic investigation of critical universality classes—particularly the Grüneisen ratio—remains absent, despite their being hallmark characteristics of quantum criticality.

\section{Holographic quantum critical point}

We consider the simple holographic model exhibiting magnetic field-induced QPTs, with the action given by
\begin{equation}\label{eq:action}
S=\frac{1}{16\pi G_N} \int d^5x \sqrt{-g}\Big( R+\frac{12}{L^2}-\frac{1}{4}F_{ab}F^{ab}+\frac{k}{24} \epsilon^{abcde}A_a F_{bc} F_{de} \Big) \,,
\end{equation}
where $G_N$ is the Newton's constant, $L$ denotes the AdS radius, and $k$ is the Chern-Simons coupling. The gauge field strength is defined as $F_{ab}=\partial_a A_b-\partial_b A_a$. For $k=2/\sqrt{3}$, the action~\eqref{eq:action} corresponds to the bosonic sector of five dimensional minimal supergravity and is a consistent truncation of Type IIB supergravity or M-theory~\cite{Buchel:2006gb,Gauntlett:2006ai}. From a bottom-up perspective, the parameter $k$ can be treated as free. Without loss of generality, we will set $16\pi G_N=L=1$.

The quantum many-body system via the holographic approach resides in three-dimensional space $\mathbb{R}^3$ with coordinates $(x_1, x_2, x_3)$, and features both a finite charge density and a magnetic field $B$ aligned with the $x_3$-axis. The magnetic field-driven QPT is holographically realized through transition between distinct infrared scaling geometries: a fractionalized phase characterized by a charged AdS$_2\times \mathbb{R}\times \mathbb{R}^2$ horizon for $B<B_c$, and a cohesive phase with a neutral AdS$_3\times \mathbb{R}^2$ horizon for $B>B_c$~\cite{DHoker:2012rlj,Zhao:2025gej}. Similar fractionalization transitions have also been identified in other holographic models of QPTs~\cite{Donos:2012js,Hartnoll:2011pp}. A notable property of this holographic framework is that $z=3$ emerges intrinsically from the underlying gravitational dynamics, in contrast to the Hertz-Millis approach where $z=3$ is derived from a one-loop approximation (see Appendix~\ref{app:hertz})~\cite{Hertz:1976zz,Millis:1993,Zhu2003UniversallyDG}. The quantum critical behavior, particularly the dynamical critical exponent $z$, is governed by the Chern–Simons coupling $k$: it follows $z=k/(1-k)$ for $1/2<k\leq3/4$, and continuously transitions to $z=3$ for $k>3/4$~\cite{DHoker:2010onp,DHoker:2010zpp}. However, despite being proposed some time ago, its thermodynamic properties---including the specific heat, magnetization, and susceptibility---remain poorly understood, and its universality class and MCE have yet to be investigated. A systematic study of these aspects is thus crucial for unlocking the full potential of this framework in describing strongly correlated systems. 

\begin{figure}[t]
\centering
\includegraphics[width=1.0\textwidth]{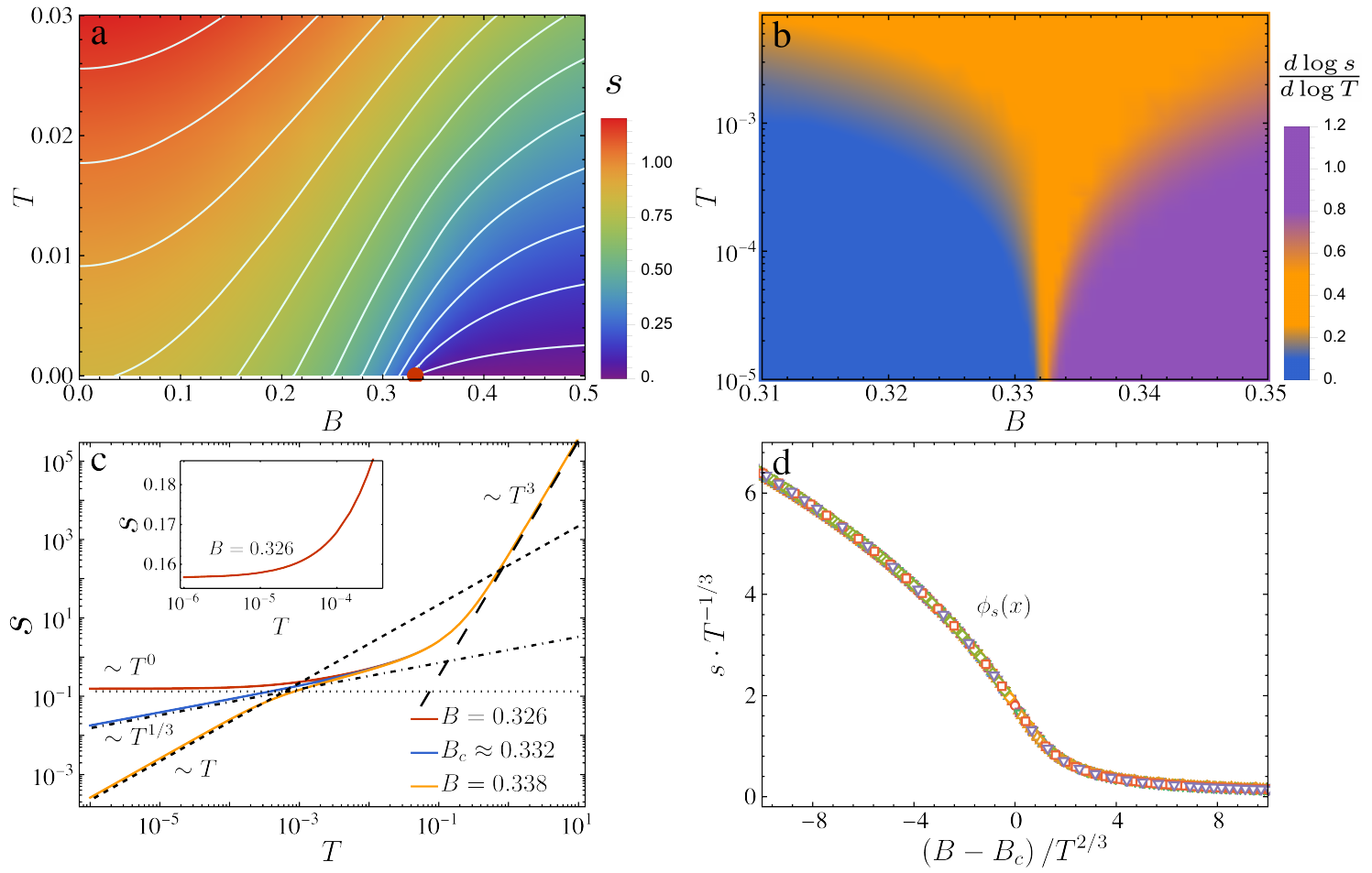}
\caption{\textbf{Entropy density $s$ and its quantum critical behavior.}
\textbf{a} Contour plot of the entropy density in $T$-$B$ plane. White lines represent the isentropic lines, while red dot represents the QCP.
\textbf{b} Temperature-dependent power $m$ of entropy $S \propto T^{m}$, where $m = \frac{d\log{s}}{d\log{T}}$. Orange region indicates the QCR hosting the quantum critical scaling $s\propto T^{1/3}$. Blue region represents the fractionalized phase, while the purple region is the cohesive phase. 
\textbf{c} Temperature dependence of entropy in different fields. Inset illustrates the zero point entropy at the fractionalized phase. 
\textbf{d} Quantum critical scaling function $\phi_s(x)$ extracted from the collapse of entropy density data for temperatures $10^{-6}\leq T\leq10^{-4}$. The plots are shown for $k=k_{\text{susy}}=2/\sqrt{3}$ and $\mu=1$.}
\label{fig:entropy}
\end{figure}
Here, we focus on the $k=2/\sqrt{3}$ supersymmetric case, corresponding to a $z=3$ QCP. We work in the grand canonical ensemble, where the free energy density depends on temperature $T$, chemical potential $\mu$, and magnetic field $B$. The equilibrium properties associated with the QPT are obtained straightforwardly using the holographic dictionary after solving the bulk equations of motion numerically. See Appendices~\ref{app:eom} and~\ref{app:num} for more details. Owing to the scale invariance of the system, all physical quantities are normalized by the chemical potential $\mu$,~\emph{i.e.} we set $\mu=1$.

In Fig.~\ref{fig:entropy}\textbf{a}, we show a contour plot of entropy density $s$ in the $T$-$B$ plane. The QCR can be identified by the power of entropy, as shown in Fig.~\ref{fig:entropy}\textbf{b}. Within a well-defined cone-like region, the entropy exhibits quantum critical scaling governed by the exponent $d/z$, where $d$ represents the effective spatial dimension of the QPT. Unlike the conventional quantum critical system, in which the isentropic lines reach their minimum in the QCR, the temperature decreases monotonically with $B$ during an adiabatic demagnetization process. Notably, this cooling effect has significant implications for the Gr\"uneisen ratio, as will be discussed in Sec.~\ref{secGR}. Fig.~\ref{fig:entropy}\textbf{c} displays the temperature dependence of $s$ for various fields. At high temperatures, these entropy curves in different fields overlap, following the same trivial $T^3$ scaling. In the cohesive phase ($B>B_c$), the low-temperature entropy crossovers to a $s \propto T$ scaling, which describes the excitation of a chiral fermion propagating along the direction of the magnetic field ($x_3$-axis)~\cite{DHoker:2012rlj}. Above the QCP at $B=B_c$, the low-temperature entropy exhibit a quantum critical behavior $s \propto T^{d/z} \propto T^{1/3}$. In the fractionalized phase ($B<B_c$), there exists a zero point entropy at the ground state, where the system hosts macroscopic degeneracy.

Near the QCP, singular part of the thermal entropy exhibits a universal behavior $s=T^{d/z} \phi_s(x)$ with $d/z=1/3$, which is obtained by the semi-analytical matching method~\cite{DHoker:2010zpp,DHoker:2010onp}. Here, $\phi_s(x)$ is the universal scaling function with $x \equiv (B-B_c)/ T^{2/3}$. At the QCP, the magnetic field effectively freezes the $x_1$ and $x_2$ directions in the near-horizon geometry, such that the resulting geometry corresponds precisely to the three-dimensional part of the ``Schr\"odinger" spacetime~\cite{Son:2008ye,Balasubramanian:2008dm,DHoker:2010onp}, indicating that the low-energy dynamics of the dual theory is reduced to an effective $1+1$ system with the spatial dimension $d=1$. Applying the standard scaling ansatz, $s\sim T^{d/z} \phi_s\left( (B-B_c) T^{-1/z\nu} \right)$, yields $z=3$ and $z\nu=3/2$, which implies $\nu = 1/2$. As shown in the Fig.~\ref{fig:entropy}\textbf{d}, the universal quantum critical scaling function $\phi_s(x)$ is obtained from the data collapse of entropy density. Remarkably, we note that a $z=3$ QCP, corresponding to a cubic dispersion relation $\omega\sim k^3$, has also been observed recently in some heavy-fermion systems~\cite{Tokiwa2009,Bin2020,Zhan2025CriticalFA}.

\begin{figure}
\centering
\includegraphics[width=0.85\textwidth]{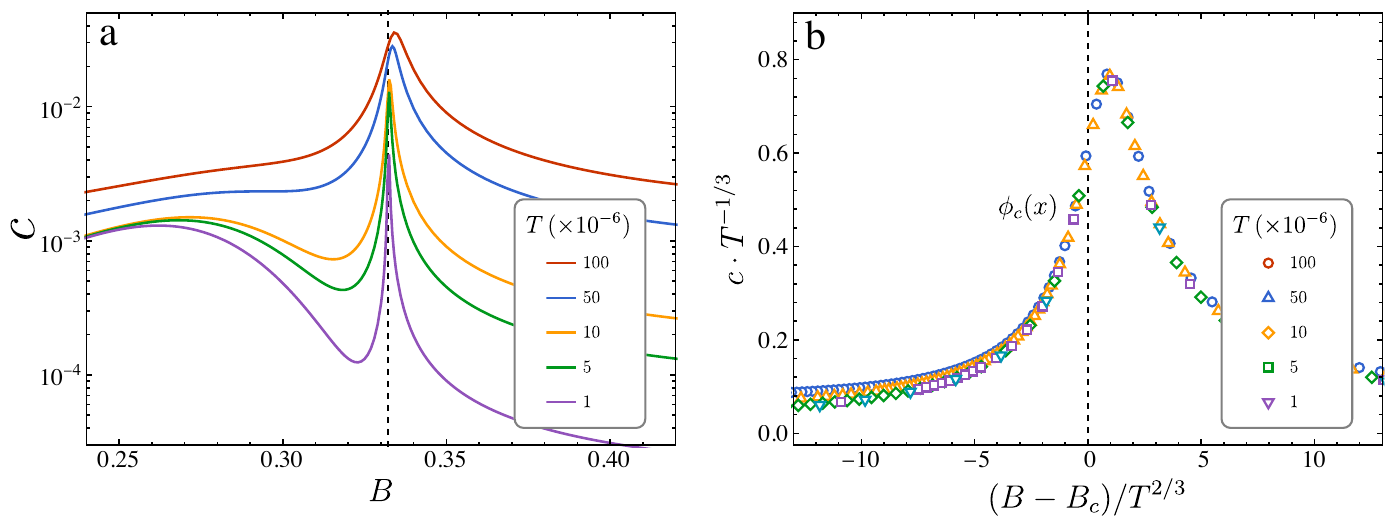}
\caption{\textbf{Specific heat near the QCP.}
\textbf{a} Isothermal specific heat as a functions of magnetic field. The vertical dashed line marks the critical field $B_c\approx 0.332$ of the QCP.
\textbf{b} Scaling function $\phi_c(x)$ of the specific heat.}
\label{fig:sheat}
\end{figure}

Furthermore, specific heat $c=T (\partial s/\partial T)_{B,\mu}$ also exhibits universal behaviors near the QCP. Fig.~\ref{fig:sheat}\textbf{a} illustrates the isothermal specific heat as a function of the magnetic field. As the temperature decreases, the specific heat density exhibits a pronounced peak in the vicinity of the critical field (vertical dashed line). As shown in Fig.~\ref{fig:sheat}\textbf{b}, the specific heat exhibits accurate data collapse onto a universal scaling function $\phi_c(x)$ near the QCR, thereby validating the critical exponents $z,\nu$ derived from entropy scaling (see Fig.~\ref{fig:entropy}\textbf{c}-\textbf{d}). Different from the conventional scaling function with double peaks~\cite{Channarayappa2024}, the sharp peaks in Fig.~\ref{fig:sheat}\textbf{a} collapse onto a single peak in Fig.~\ref{fig:sheat}\textbf{b}. 
Thus, only the right peak of $c$ reliably identifies the boundary of the QCR $B-B_c \propto T^{1/z\nu}$; the left boundary cannot be determined from the specific heat. We attempted to identify the left QCR boundary using other physical quantities, yet no suitable quantity was found. Consequently, we define the full QCR by symmetrically reflecting the right QCR boundary.

\section{Magnetization, order parameter and universality class of the QCP}
\begin{figure} 
\centering
\includegraphics[width=0.85\textwidth]{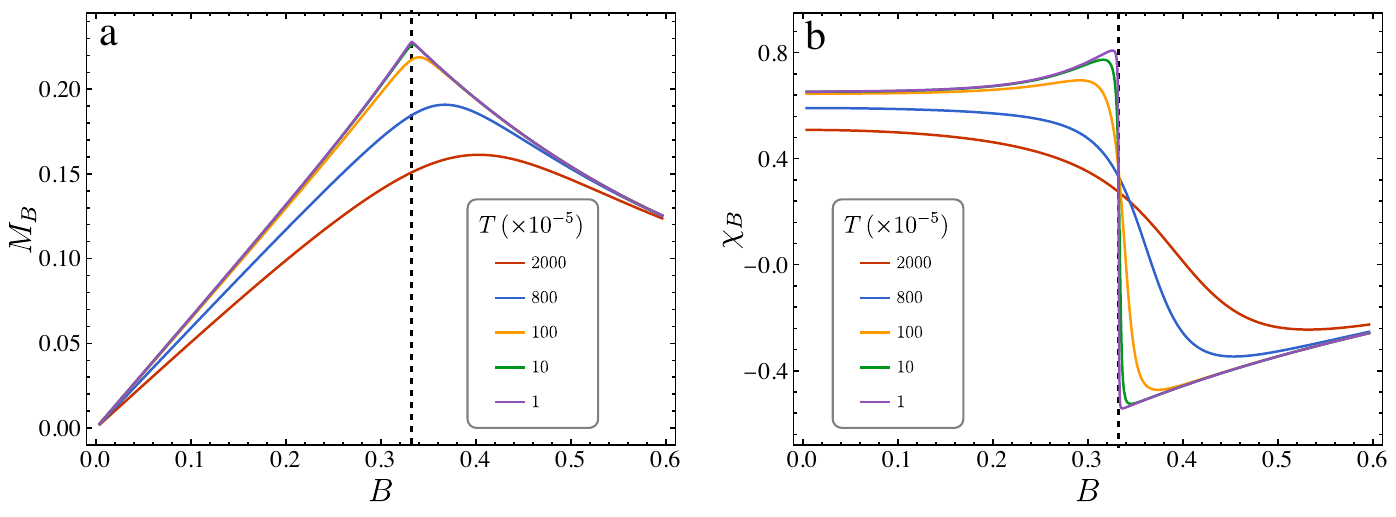} %
\caption{\textbf{Magnetic properties of the QPT.}
\textbf{a} Magnetic field dependence of magnetization $M_B$ at different temperatures. The vertical dashed line marks the location of QCP.
\textbf{b} Behavior of magnetic susceptibility $\chi_B$ with respect to $B$ for various temperatures. The plots are shown for $k=k_{\text{susy}}=2/\sqrt{3}$ and $\mu=1$.}
\label{fig:mBchi}
\end{figure}
The magnetization $M_B=-(\partial w/\partial B)_{T,\mu}$, derived from the free energy density $w$, is shown in Fig.~\ref{fig:mBchi}\textbf{a} as a function of the magnetic field $B$ at different temperatures. For each curve, $M_B$ exhibits a non-monotonic behavior: it increases from zero field, reaches a maximum at an intermediate value of $B$, and then decreases upon the increasing of $B$. While the magnetization varies smoothly at high temperatures, its derivative becomes discontinuity at low temperatures, signaling a cusp singularity across the QPT (see Fig.~\ref{fig:mBchi}\textbf{a}). From the maximum of $M_B$ at each temperature $T_0$, located at $B_0$ in Fig.~\ref{fig:mBchi}\textbf{a}, we find that they obey the scaling relation $B_0-B_c\propto T_0^{0.69}$. The extracted exponent $0.69$ is close to the expected value $1/z\nu=2/3$ for quantum critical crossovers. This suggests that $M_B$, like the specific heat, may be used to identify the right boundary of the QCR. Notably, since no symmetry change is associated with this QPT, it was proposed in~\cite{DHoker:2010zpp} that this is a metamagnetic QPT, which is empirically characterized by a rapid increase in magnetization toward a critical field, followed by saturation at large fields~\cite{Millis2001MetamagneticQC}. In contrast, our explicit calculations reveal no such rapid magnetization increase; instead, a cusp singularity emerges near the QCP (see Fig.~\ref{fig:mBchi}\textbf{a}), indicating a departure from the conventional picture of metamagnetic QPT~\cite{Millis2001MetamagneticQC}.

Fig.~\ref{fig:mBchi}\textbf{b} displays the magnetic susceptibility $\chi_B=(\partial M_B/\partial B)_{T,\mu}$ as a function of $B$. At high temperatures, $\chi_B$ decreases monotonically form a positive value to a negative value with the increasing of $B$. At low temperatures, it first increases to a maximum before reaching $B_c$, then decreases to a minimum after passing $B_c$, and finally increases again with further increase in $B$. Notably, $\chi_B$ becomes discontinuous in the zero temperature limit. The distinct behavior of both $M_B$ and $\chi_B$ lead us to conclude that the dual strongly coupled quantum many-body system undergoes a second order QPT from a paramagnetic phase ($\chi_B>0$) for $B<B_c$ to a diamagnetic phase ($\chi_B<0$) for $B>B_c$. Approaching the QCP, the magnetization varies linearly with the magnetic field, while the susceptibility remains approximately constant.

\begin{figure} 
\centering
\includegraphics[width=0.85\textwidth]{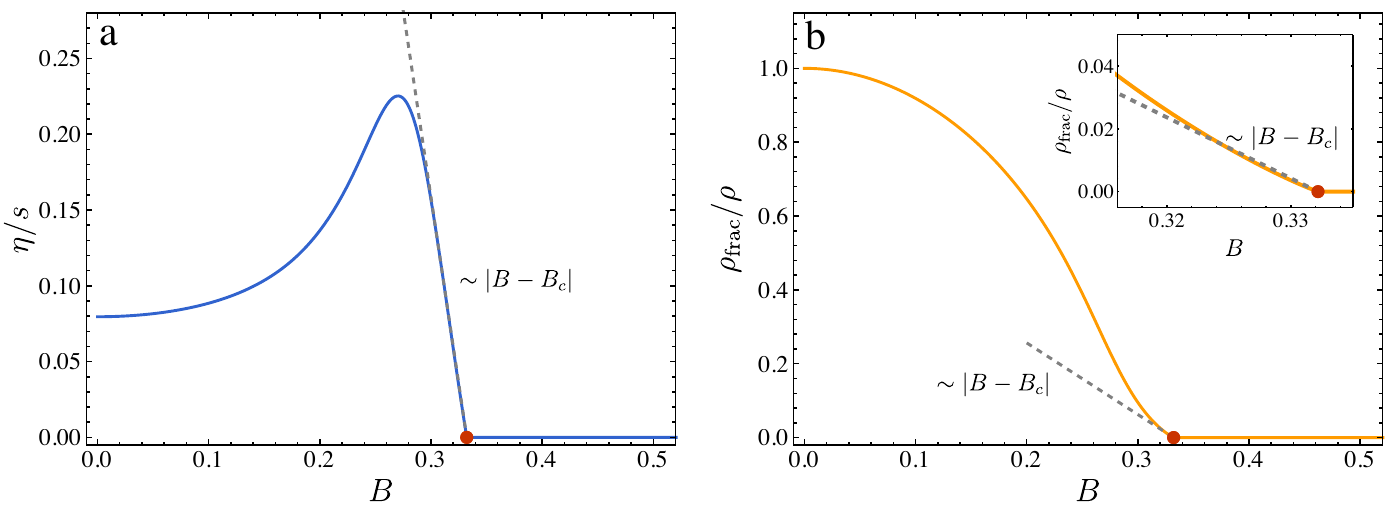} 
\caption{\textbf{Order parameter of the holographic QCP.}
\textbf{a} Magnetic field dependence of the ratio of parallel shear viscosity to entropy density $\eta/s$ at extremely low temperature $T=10^{-6}$.
\textbf{b} The ratio of fractionalized charge to total charge $\rho_{\text{frac}}/\rho$ at extremely low temperature $T=10^{-6}$. Inset illustrates the behavior of $\rho_{\text{frac}}/\rho$ near the QCP. The plots are shown for $k=k_{\text{susy}}=2/\sqrt{3}$ and $\mu=1$.}
\label{fig:op}
\end{figure}
Before, the critical exponents $z$ and $\nu$ for the QPT have been extracted from the scaling behavior of the entropy density, with their validity further corroborated by the data collapse of both the entropy and the specific heat. A complete characterization of the QCP requires determining the full set of critical exponents. The QPT considered here goes beyond the conventional Landau-Ginzburg paradigm, as it occurs without any symmetry breaking. In such cases, suitable order parameters are often defined in the literature by taking the zero temperature limit of a finite temperature quantity that is finite in one phase but vanishes in the other. Here, we identify two physical quantities that serve as suitable order parameters: the ratio of parallel shear viscosity to entropy density $\eta/s$~\cite{Zhao:2025gej}, and the ratio of fractionalized charge to total charge, $\rho_{\text{frac}}/\rho$. The fractionalized charge $\rho_{\text{frac}}$ denotes the contribution to the electric flux originating from the charged black hole horizon. Particularly, the emergence of charge fractionalization may hold the key to understanding the Fermi surface reconstruction that accompanies a second order QPT. As shown in Fig.~\ref{fig:op}, both $\eta/s$ and $\rho_{\text{frac}}/\rho$ are finite for $B<B_c$, decrease linearly with $B$ near the QCP, and vanish identically for $B>B_c$.

According to the scaling theory for a second order QPT, this linear dependence on $B-B_c$ implies that the critical exponents for the order parameter and susceptibility (defined as the derivative of the order parameter with respect to the magnetic field) are given by $\beta=1$ and $\gamma=0$, respectively. Employing the standard relations between critical exponents~\cite{sachdev2015,Continentino_2017}, we then derive the remaining critical exponents, including specific heat exponent $\alpha$, the critical isotherm exponent $\delta$ and the anomalous dimension $\eta$. The complete critical exponents is summarized in Table~\ref{table}. To the best of our knowledge, this set of critical exponents does not correspond to any previously known universality class. To investigate whether the properties of this $z=3$ QCP are generic, we also examine the case $k=3/2$, which yields the same dynamical critical exponent. We find that the value of full critical exponents remain unchanged, and in particular, the universal scaling function $\phi_s(x)$ for the entropy takes an identical form (see Appendix~\ref{app:pdz3}). Given that this universality class with $z=3$ is derived from the holographic framework involving a charged black hole with Chern-Simons interactions, we designate it the ``EMCS cubic universality class." 
\begin{table}[htbp]
\centering 
\renewcommand{\arraystretch}{1.8}
\setlength{\tabcolsep}{3.6mm}
\begin{tabular}{|c|c|c|c|c|c|c|c|} 
\hline Critical exponent & $\alpha$ & $\beta$  & $\gamma$  & $\delta$ & $\eta$ & $\nu$ & $z$ \\ 
\hline  EMCS theory & 0    &  1  &    0  & 1 &  2 & $\frac{1}{2}$ & 3 \\
\hline 
\end{tabular} 
\caption{Critical exponents of the EMCS cubic QCP, which lie beyond any known universality class.} \label{table}
\end{table}

\section{Universally diverging Gr\"uneisen ratio} \label{secGR}
\begin{figure} 
\centering
\includegraphics[width=0.85\textwidth]{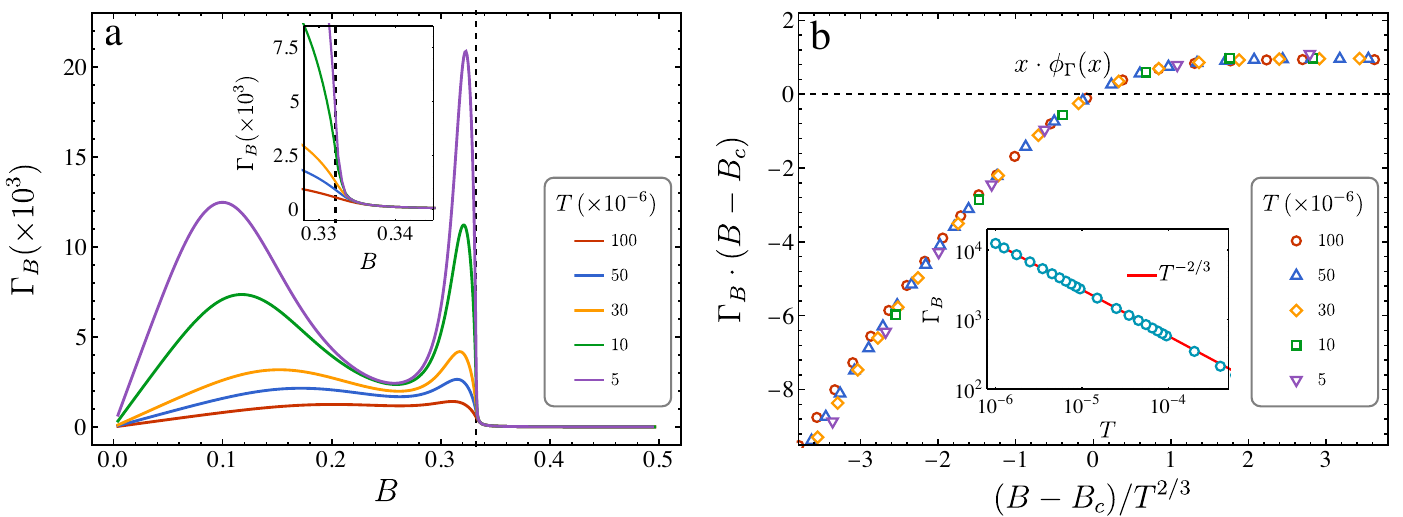} 
\caption{\textbf{Universally diverging Gr\"uneisen ratio near the QCP}.
\textbf{a} Magnetic Gr\"uneisen ratio $\Gamma_B$ as a function of magnetic field $B$. Inset illustrates the Gr\"uneisen ratio near the QCP. 
\textbf{b} Scaling function $\phi_{\Gamma}(x)$ of Gr\"uneisen ratio near the QCP. The inset shows the temperature-dependent scaling of $\Gamma_B$ at $B=B_c$. The plots are shown for $k=k_{\text{susy}}=2/\sqrt{3}$ and $\mu=1$.}
\label{fig:gruneisen} 
\end{figure}
Here, we explore a novel phenomenon ---the magnetocaloric effect--- of this holographic QCP~\cite{Zhu2003UniversallyDG}. Magnetic Gr\"uneisen ratio $\Gamma_B= \frac{1}{T}\left(\frac{\partial T}{\partial B}\right)_{s,\mu} =-\frac{(\partial s/\partial B)_{T,\mu}}{T(\partial s/\partial T)_{B,\mu}}$, defined as the temperature variation during an adiabatic demagnetization process, can character the MCE.  Fig.~\ref{fig:gruneisen}\textbf{a} shows the magnetic Gr\"uneisen ratio $\Gamma_B$ as a function of magnetic field $B$ for various temperatures. At high temperatures, $\Gamma_B$ initially increases with the magnetic field, forming a peak before reaching $B_c$, and then decreases monotonically. This peak grows and ultimately diverges as the temperature is lowered. Notably, the sign of $\Gamma_B$ remains positive across all magnetic fields, which is consistent with the monotonic increase of temperature with $B$ along each isentropic curve, as shown in Fig.~\ref{fig:entropy}\textbf{a}.

In particular, $\Gamma_B$ displays only a single peak across the QCP, in sharp contrast with the conventional peak-dip structure and sign reversal in Gr\"uneisen ratio seen in most QPTs~\cite{Garst2005SignCO}. It is empirically observed --- and confirmed in many models --- that the entropy is maximal at the QCP, meaning it decreases more slowly with temperature along $B=B_c$ (or $x=0$) than in other phases. Theoretically, however, there could exist quantum phases in which $s$ decreases even more slowly than that along $B=B_c$. In such a case, the isentropes would vary monotonically with the external field, and the Gr\"uneisen ratio would exhibit no sign change. Interestingly, our model indeed realizes a concrete example of this type, exhibiting no sign reversal in the Gr\"uneisen ratio $\Gamma_B$. Notably, similar phenomena in $\Gamma_P$ were identified in the 3D XY model of pressure induced QPT in recent studies~\cite{2025arXiv250917362Z}. For practical implementations such as magnetic refrigeration, the absence of a sign reversal in the Gr\"uneisen parameter $\Gamma_B$ provides a significant advantage: the cooling procedure is not constrained by the need to operate either above or below the critical magnetic field.

To better understand the quantum critical MCE in this model, we perform a scaling analysis of the diverging Gr\"uneisen ratio $\Gamma_B$. As shown in Fig.~\ref{fig:gruneisen}\textbf{b}, we show the data collapse of $\Gamma_B$ near the QCP. The $\Gamma_B (B-B_c)$, when plotted against $x$, collapse onto universal curves, confirming the quantum critical scaling of the Gr\"uneisen ratio. In the insert of Fig.~\ref{fig:gruneisen}\textbf{b}, we plots the temperature dependence of $\Gamma_B$ above the QCP ($B=B_c$ or $x=0$). Above the QCP, the numerical results follows the scaling $\Gamma_B\propto T^{-1/z\nu} = T^{-2/3}$, demonstrating the universally diverging of Gr\"uneisen ratio~\cite{Zhu2003UniversallyDG}. On either side of the $B=B_c$, the $T^{-2/3}$ scaling only persists in an intermediate temperature range and breaks down as the system passing through the QCR at extremely low temperatures. Derived from critical scaling of the entropy density $s=T^{1/3}\phi_s(x)$, the Gr\"uneisen ratio in the QCR should satisfy a scaling form of $\Gamma_B =T^{-2/3} \phi_\Gamma(x)$, where $\phi_\Gamma(x)$ is the universal scaling function of $\Gamma_B$. 

\section{Discussions}
We have discovered ``EMCS cubic universality class'' for the first time and investigated associated magnetocaloric effect in strongly coupled systems hosting a magnetic field-driven QCP, described holographically by a five-dimensional EMCS theory. The presence of quantum criticality are demonstrated through the scaling behavior of the entropy and the specific heat, as shown in Fig.~\ref{fig:entropy} and Fig.~\ref{fig:sheat}, respectively. The magnetic properties of the system are studied in Fig.~\ref{fig:mBchi}. An interesting finding is the emergence of a cusp singularity in the magnetization $M_B$ across the transition---a feature distinct from conventional metamagnetic QPTs. The behavior of magnetic susceptibility $\chi_B$ further indicates a continuous second order QPT from a paramagnetic state to a diamagnetic phase. As shown in Fig.~\ref{fig:op}, the order parameters of the QPT are identified as the ratio of parallel shear viscosity to entropy density $\eta/s$, and the ratio of fractionalized charge to total charge $\rho_{\text{frac}}/\rho$.

More importantly, our work provides the first demonstration of the universal divergence of the Gr\"uneisen ratio near a holographic QCP. Remarkably, in our system the magnetic Gr\"uneisen ratio $\Gamma_B$ exhibits neither the sign reversal nor the peak-dip structure commonly observed in many quantum critical systems (see Fig.~\ref{fig:gruneisen}). Above the QCP, we reveal the quantum critical scaling $\Gamma_B \propto T^{-1/z\nu} \propto T^{-2/3}$ with $z=3$ and $\nu=1/2$. These results confirm the utility of the magnetic Gr\"uneisen ratio as a precise tool for locating the QCP and determining critical exponents in strongly correlated holographic quantum matter.

In addition to the $z=3$ QCP discussed above, our model is capable of realizing other dynamical exponents. As a demonstration, we present in Appendix~\ref{app:z2} a $z=2$ QCP exhibiting a third order QPT. Due to its third order nature, a complete set of critical exponents cannot be determined via the standard relations between critical exponents~\cite{sachdev2015,Continentino_2017}. Nevertheless, we find that the Gr\"uneisen ratio still diverges, and provides an effective way to extract $z\nu$. Remarkably, the value of $z\nu$ obtained from $\Gamma_B$ is consistent with that deduced from the data collapse of the entropy and specific heat. This remarkable consistency suggests that the universal scaling laws governing the divergence of the Gr\"uneisen ratio, established for second order QPTs~\cite{Zhu2003UniversallyDG,Garst2005SignCO}, could be extended to third-order transitions.

Our work provides a novel platform for investigating exotic metallic quantum criticality in strongly correlated systems such as heavy-fermion materials, offering new insights into field-induced QPTs and charting promising directions for future research. Furthermore, the holographic construction naturally yields a class of $z=3$ QCPs, which bears a promising connection to the cubic dispersion relation recently observed in certain heavy-fermion systems such as~CeRh$_6$Ge$_4$. We have demonstrated that a remarkably simple holographic model can capture rich quantum critical behaviors reminiscent of heavy-fermion systems. Notably, it offers a computationally convenient framework in regimes where traditional methods currently face difficulties. Given that critical behaviors are largely insensitive to the microscopic details of the theory, the holographic approach proves to be a powerful tool for uncovering different universality classes and scaling functions. A natural extension, incorporating additional degrees of freedom, would enable the model to capture other experimentally relevant phenomena like electrical conductivity, paving the way toward a holographic description of strange metal behavior consistent with observations~\cite{Zhan2025CriticalFA}.

It would be important to compute the fermion spectral function following~\cite{Liu:2009dm,Cubrovic:2009ye,Faulkner:2009wj}, thereby enabling direct access to collective excitations near this QCP. The fractionalization picture described here may be an important element for understanding the large-to-small Fermi surface transition, shedding new light on the origin of strange metality in heavy-fermion materials~\cite{Senthil:2002vtt,Hartnoll:2011pp}. In the present model, numerical evidence points to a finite entropy density in the ground state of the fractionalized phase. To circumvent this issue and better model realistic quantum critical systems, future work could focus on setups that exhibit vanishing ground state entropy in both phases. One promising route is to introduce helical order, which supports a spatially modulated phase spanning the QCP~\cite{Ammon:2016szz,Nakamura:2009tf}. Such a configuration may be viewed as a holographic realization of a chiral magnetic spiral~\cite{Basar:2010zd} and could be closely related to nematic phases observed in metamagnetic materials~\cite{Fradkin:2010ARCMP}.

%

\ack{We thank Ning Xi and Xiang-Xi Zeng for useful discussions. This work is supported by the National Natural Science Foundation of China (grant No.\,12525503, No.\,12505087, No.\,12588101, and No.\,12447101), and the Strategic Priority Research Program of Chinese Academy of Sciences (Grant No.~XDB1270100). We acknowledge the use of the High Performance Cluster at the Institute of Theoretical Physics, Chinese Academy
of Sciences.}





\appendix

\section{Hertz-Millis theory}\label{app:hertz}
Here, we present a brief introduction of Hertz-Millis theory and discuss key similarities and differences with our holographic model. The Hertz-Millis theory of quantum criticality provides a standard framework for describing magnetic field-induced QPTs~\cite{Hertz:1976zz,Millis:1993}. In this approach, the fermionic degrees of freedom are integrated out, yielding an effective bosonic theory for the order parameter $\phi$ that can then be studied using renormalization group methods. The effective action in $d$ spatial dimensions takes the form
\begin{equation}
S_{HM}= \sum_{\boldsymbol{k}, i\omega_n} \left( \delta +\boldsymbol{k}^2 +\frac{|\omega_n|}{\Gamma_k} \right) |\phi_{\boldsymbol{k}, i\omega_n}|^2 +u\int_0^{1/T} d\tau \int d^d \boldsymbol{r} \phi(\boldsymbol{r},\tau)^4 \,,
\end{equation}
where $\Gamma_k=\Gamma_0 \boldsymbol{k}^{z-2}$ is the Landau damping rate, and $\delta$ and $u$ are parameters inherited from the underlying electron Hamiltonian. In particular, $\delta$ serves as the tuning parameter that changes sign as one varies some control parameter in the microscopic model, while $u$ denotes the strength of the interaction, which is assumed to depend negligibly on Hamiltonian parameters over the range of interest. The sum runs over bosonic Matsubara frequencies $\omega_n$ and wave vectors $\boldsymbol{k}$. Here, the scalar field $\phi$ represents the order parameter of the QPT. The case $z=2$ describes an antiferromagnetic spin-density-wave transition in a metal, whereas $z=3$ may apply to the critical endpoint of a metamagnetic first-order transition in $d=2,3$. This approach, however, relies on the assumption that the fermions degrees of freedom can be safely integrated out---a procedure that is justified only when they are treated as high energy modes~\cite{Lohneysen:2007zz,Metlitski:2010pd}.

In $d=1$, the Hertz-Millis action is invariant under the scaling transformation $\omega\to \lambda^3 \omega \,, k\to \lambda k$ at the QCP if $z=3$. Moreover, $B-B_c$ has scaling dimension $2$ and acts as a relevant coupling in the theory. These properties are consistent with the scaling behavior $s\propto T^{1/3}$ for Chern Simons coupling $k>3/4$.

\section{Equations of motion and holographic renormalization}\label{app:eom}
The equations of motion derived from the EMCS action~\eqref{eq:action} are given by
\begin{equation}
\begin{split}  \label{eq:eom}
R_{ab}-\frac{1}{2}g_{ab}\left( R+12 \right)-\frac{1}{2}\left(F_{ac}F_b^{\ c}-\frac{1}{4}g_{ab}F_{cd}F^{cd} \right) =&0 \,,   \\ 
\nabla_{b}F^{ba} +\frac{k}{8} \epsilon^{abcde}F_{bc}F_{de} =&0 \,.
\end{split}
\end{equation}
The dual magnetic field-induced QPT is described by the following dyonic black hole~\cite{DHoker:2010onp,Zhao:2025gej}
\begin{equation} \label{eq:ansata}
\begin{split}
ds^2&=\frac{1}{r^2}\Big[- \big(f-h^2p^2\big)dt^2+2ph^2 dtdx_3+g\big( dx_1^2+dx_2^2 \big)+h^2 dx_3^2+\frac{dr^2}{f}\Big]\,,  \\
A&=A_t dt-\frac{B}{2}x_2 dx_1+\frac{B}{2}x_1 dx_2-A_z dx_3 \,,
\end{split}
\end{equation}
where $f, g, h, p, A_t,$ and $A_z$ are functions of radial coordinate $r$ which is identified as the energy scale of the dual boundary theory. In these coordinates, the asymptotically AdS boundary lies at $r=0$, and the black hole event horizon is located at $r=r_h$, where all bulk fields are regular with $f(r_h)=0$ (See Fig.~\ref{fig:adscft}). The constant $B=F_{x_1x_2}$ denotes the background magnetic field, which is aligned along the $x_3$ direction.

Inserting the ansatz~\eqref{eq:ansata} into the above field equations~\eqref{eq:eom} yields a set of six coupled second order differential equations:
\begin{equation}
\begin{split}
\Big[\frac{g h}{r} \Big(A_t'+p A_z' \Big) \Big]' +k B A_z' &=0 \,,\\
\Big[\frac{g}{r h} \Big(p h^2A_t'-(f-h^2p^2)A_z' \Big) \Big]' -k B A_t' &=0\,, \\
\frac{f''}{f} +\frac{h''}{h} +\left( \frac{f'}{f} -\frac{g'}{g} \right)\left( \frac{h'}{h} -\frac{3}{r}\right) -\frac{g'^2}{2g^2} -\frac{h^2p'^2}{f} +\frac{r^2A_z'^2}{2h^2} -\frac{B^2r^2}{fg^2} -\frac{r^2(A_t'+pA_z')^2}{f} &=0 \,,\\
\frac{g''}{g}+\frac{g'}{3g}\left(\frac{f'}{f}-\frac{g'}{g}-\frac{h'}{h}-\frac{3}{r}\right)+\frac{2h'}{h}\left(\frac{1}{r}-\frac{f'}{3f}\right)+\frac{B^2 r^2}{3f g^2}-\frac{h^2p'^2}{3f}&=0 \,, \\
\frac{h''}{h}+\left(\frac{h'}{h}-\frac{g'}{2g}\right)\left(\frac{2f'}{3f}+\frac{g'}{3g}-\frac{2}{r}\right)+\frac{h^2p'^2}{3f}+\frac{r^2A_z'^2}{2h^2}-\frac{B^2r^2}{3f g^2}&=0 \,, \\
p''+\left( \frac{g'}{g}+\frac{3h'}{h}-\frac{3}{r}\right)p'-\frac{r^2 A_t' A_z'}{p h^2}-\frac{r^2 A_z'^2}{h^2}&=0 \,, 
\end{split}
\end{equation}
where the prime represents the derivative with respect to $r$. Additionally, the following constraint equation:
\begin{equation} \label{constraint}
\left(\frac{g'}{g} +\frac{f'}{2f}\right)\left( \frac{g'}{g} +\frac{h'}{h} -\frac{3}{r} \right) -\frac{3g'^2}{4g^2} -\frac{3h'}{rh} +\frac{6(f-1)}{r^2 f} +\frac{Br^2}{4fg^2} -\frac{r^2A_z'^2}{4h^2} +\frac{h^2p'^2}{4f} +\frac{r^2(A_t+pA_z')^2}{4f}=0\,,
\end{equation}
is automatically satisfied when the above equations hold. Denoting the constraint equation~\eqref{constraint} as $\text{CON}(r)=0$, we use it as a check of numerical accuracy, as demonstrated in Appendix~\ref{app:num}.

The asymptotic expansion of the bulk fields near the AdS boundary $r\to0$ are given by
\begin{equation}
\begin{split} \label{uvexpansion}
f(r)&= 1+\frac{B^2}{6}r^4\ln{r}+f_4 r^4+\cdots \,, \quad 
g(r)= 1-\frac{B^2}{12}r^4\ln{r}-h_4 r^4+\cdots  \,, \\
h(r)&= 1+\frac{B^2}{12}r^4\ln{r}+h_4 r^4+\cdots  \,, \quad
p(r)= p_4 r^4+\cdots    \,, \\
A_t(r)&= \mu +A_{t2} r^2+\cdots  \,, \quad\qquad
A_z(r)= A_{z2} r^2 +\cdots\,,
\end{split}
\end{equation}
where the reparameterization freedom $r\to r+\xi$ in the radial direction has been fixed by setting $f'(r)|_{r=0}=0$. Near the black hole horizon $r=r_h$, the bulk fields are regular and can be expanded as
\begin{equation}
\begin{split}
f(r)&= (r_h-r)f_0+\cdots  \,,\quad g(r)= g_0+\cdots  \,,\qquad\qquad\qquad  h(r)= h_0+\cdots  \,, \\
p(r)&= (r_h-r)p_0+\cdots    \,,\quad A_t(r)= (r_h-r)A_{t0}+\cdots \,,\quad A_z(r)= A_{z0}+\cdots \,.  
\end{split}
\end{equation}

Note that, the full bulk solutions for non-vanishing magnetic field are only accessible numerically. We therefore solve the background equations of motion via the pseudo-spectral method~\cite{boyd2001chebyshev}, implementing appropriate asymptotic and horizon conditions. A useful scaling symmetry 
\begin{equation}
 \left(t,x_1,x_2,x_3,r \right)\to y\left( t,x_1,x_2,x_3,r \right),\, \left(A_t, A_z \right)\to \left( \frac{A_t}{y}, \frac{A_z}{y} \right),\, B\to \frac{B}{y^2} ,\, (f,g,h,p)\to(f,g,h,p)\,,
\end{equation}
allows us to fix the event horizon to $r_h=1$ in the numerical computations. The numerical solutions are validated by the constraint equations~\eqref{constraint}. Further numerical details are provided in Appendix~\ref{app:num}.

Thermodynamic quantities are straightforwardly obtained using the standard holographic dictionary after solving the bulk equations of motion. Here, we collect the relevant key results (see~\cite{Cai:2024tyv} for a complete derivation). In particular, the energy density $\epsilon$, the anisotropic transverse/longitudinal pressures $\mathcal{P}_{\perp/\parallel}$, and the charge density $\rho$ are expressed as
\begin{equation}
\epsilon=-3f_4\,,\qquad    \mathcal{P}_\perp=-\frac{B^2}{4}-f_4-4h_4\,, \qquad 
\mathcal{P}_\parallel  =-f_4+8h_4 \,, \qquad \rho= -2A_{t2} \,.
\end{equation}
The temperature $T$ and entropy density $s$ of the boundary system are determined by the surface gravity and the Bekenstein–Hawking entropy, respectively
\begin{equation} \label{eq:Ts}
T=-\frac{f'(r)}{4\pi}\Big|_{r=r_h} \,,\qquad  s=\frac{4\pi g(r) h(r)}{r^3} \Big|_{r=r_h}  \,.
\end{equation}
The free energy density $w$ is given by
\begin{equation}
w = -\mathcal{P}_\parallel =-\mathcal{P}_\perp -M_BB =\epsilon -Ts -\mu \rho\,,
\end{equation}
and the first law of thermodynamics takes the form
\begin{equation} \label{first}
\delta w=-s \delta T - \rho \delta\mu -M_B\delta B \,,
\end{equation}
where the magnetization $M_B$ reads
\begin{equation}
M_B = -\left( \int_0^{r_h} \left[ \frac{B}{r} \left( \frac{h}{g}-1 \right) +\frac{k}{2}(A_t A_z' - A_t' A_z) \right] dr +B\ln{r_h} \right) \,.
\end{equation}
We have verified these thermodynamic quantities extracted from our numerical calculations and confirmed that the first law~\eqref{first} is satisfied within a relative error of less than $<10^{-3}$.

\section{Numerical details}\label{app:num}

\begin{figure} 
\centering
\includegraphics[width=0.95\textwidth]{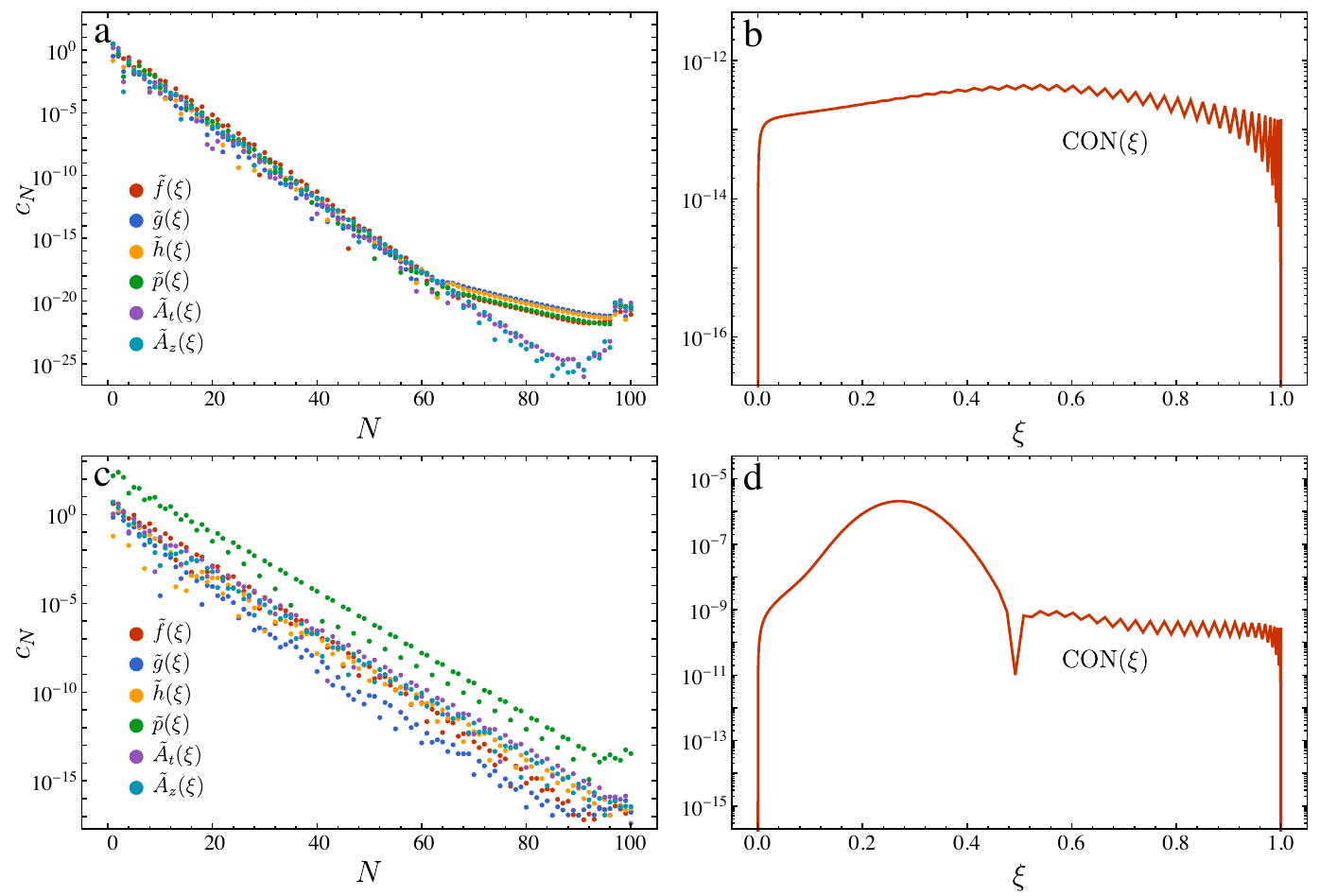}
\caption{The Chebychev coefficients $c_N$ in the expansions of bulk fields and the violation of the constraint equation for $B=0.338$ at $T=10^{-2}$(top) and $T=10^{-5}$ (bottom). The plots are shown for $k=k_{\text{susy}}=2/\sqrt{3}$ and $\mu=1$.}
\label{fig:num}
\end{figure}
In this section, we provide details on the pseudo-spectral method~\cite{boyd2001chebyshev} employed in our numerical calculations. Spectral methods typically exhibit exponential convergence with increasing grid points when the solutions are analytic. However, the presence of logarithmic terms in the boundary expansion (see~\eqref{uvexpansion}) for bulk fields spoil this rapid convergence. Additionally, studying the physics of the EMCS theory within the QCR necessitates working at extremely low temperatures, where the bulk fields develop prohibitively steep gradients near the horizon. Such numerical challenges typically require a substantial increase in grid points to maintain accuracy.

To overcome the numerical difficulties caused by the logarithmic terms in the boundary expansion and resolve the strong steep gradients, we introduce auxiliary fields utilizing the asymptotic behavior near the conformal boundary
\begin{equation}
\begin{split}
f(r)&= 1+\frac{B^2}{6}r^4\ln{r}-r^4+(1-r)r^4\tilde{f}(r) \,,\qquad g(r)= 1-\frac{B^2}{12}r^4\ln{r}+r^4\tilde{g}(r)  \,, \\
h(r)&= 1+\frac{B^2}{12}r^4\ln{r}+r^4\tilde{h}(r) \,, \qquad\qquad
p(r)= (1-r)r^4\tilde{p}(r)   \,, \\
A_t(r)&= (1-r)\left(\mu+ r\tilde{A_t}(r) \right) \,,\qquad\qquad
A_z(r)= r^2\tilde{A_z}(r)        \,.
\end{split}
\end{equation}
We then employ an analytical mesh refinement strategy by mapping the radial coordinate $r\in [0,1]$ to $\xi\in[0,1]$ via the transformation
\begin{equation}
r= 1-\frac{\sinh[\lambda(1-\xi^2)]}{\sinh\lambda} \,.
\end{equation}
Here, the the parameter $\lambda$ can gradually increase the grid points near the horizon $r=1$, which is choose to be $\lambda=|\ln{T}|-1/2$ in the numerics. Note that, this type of coordinate mapping is used in~\cite{Ammon:2016szz}. As a demonstration of numerical performance, Fig.~\ref{fig:num} shows the convergence of the Chebyshev coefficients $c_N$ and the violation of the constraint equations~\eqref{constraint}. The Chebyshev coefficients exhibit clear exponential decay, while the violation of the constraint equations remains at high precision, confirming the good performance of our numerical method.

\section{Phase diagram and the universality of the $z=3$ QCP}\label{app:pdz3}
\begin{figure}[h] 
\begin{center} 
\includegraphics[width=0.85\textwidth]{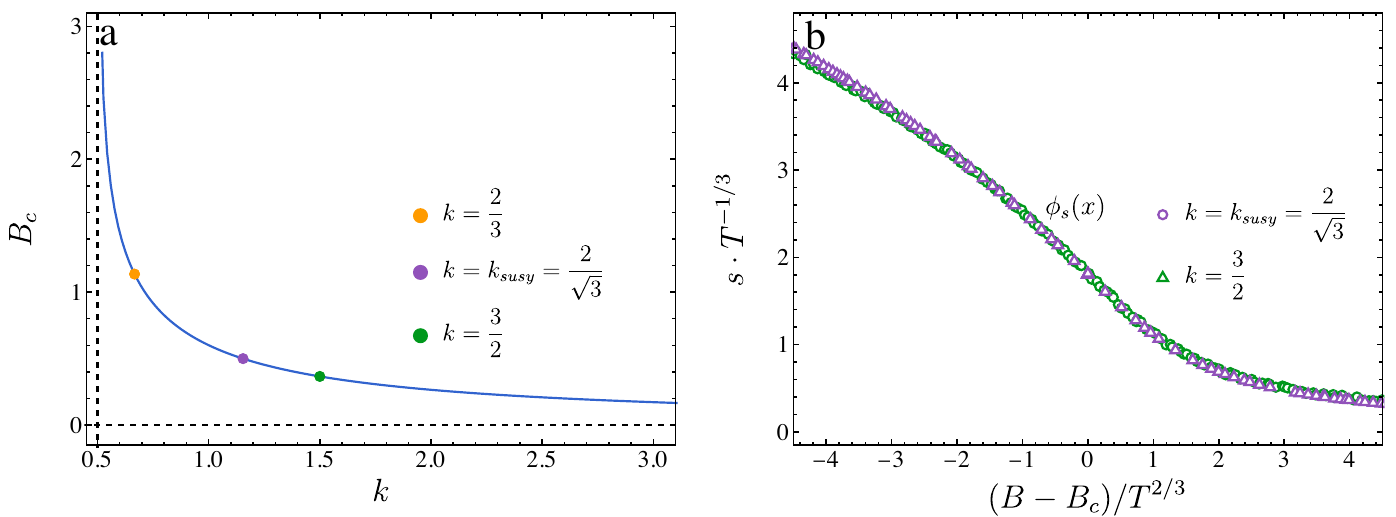}
\end{center}
\vspace{-0.3cm}
\caption{\textbf{Phase diagram and the quantum critical scaling function for $z=3$ QCP.}
\textbf{a} The critical magnetic field $B_c$ as a function of Chern-Simons coupling $k$.
\textbf{b} Comparison of the quantum critical scaling function $\phi_s(x)$ for $k=k_{\text{susy}}=2/\sqrt{3}$ and $k=3/2$.}\label{fig:pdz3}
\end{figure}

At zero temperature, the system is characterized by $B$. As this parameter varies, the near horizon geometry undergoes distinct configurations, resulting in a field-induced QPT. The zero temperature ground state, particularly the near horizon geometry at and above the QCP, is revealed in~\cite{DHoker:2010onp}. Notably, the critical magnetic field of the QPT is uniquely determined by the Chern-Simons coupling $k$, as illustrated in Fig.~\ref{fig:pdz3}\textbf{a}. We find that the critical magnetic field $B_c$ diverges near $k=1/2$ and decreases monotonically with increasing $k$. In particular, for the Chern Simons couplings considered in this work---$k=2/3$, $k=k_{\text{susy}}=2/\sqrt{3}$, and $k=3/2$---the critical magnetic fields for the QPT are determined to be $B_c \approx 1.30165$, $0.33218$, and $0.21993$, respectively. It should be noted that the ground state solution diverges at $k=1/2$ and does not represent the zero temperature limit of a black hole for $k<1/2$~\cite{DHoker:2010onp}.

Note that at $k=k_{\text{susy}}=2/\sqrt{3}$ and $k=3/2$, the dynamical critical exponent takes the same value $z=3$, and in both cases the singular part of the thermal entropy near the QCP follows the scaling form $s\sim T^{1/3}\phi_s(x)$~\cite{DHoker:2010zpp,DHoker:2010onp}. By appropriately rescaling the $x$ and the scaling function $\phi_s$, we find that the quantum critical scaling functions for these two different Chern-Simons coupling fall onto a single curve, as shown in Fig.~\ref{fig:pdz3}\textbf{b}. This demonstrates that the $z=3$ QCPs in our model are universal across different values of Chern-Simons coupling $k$ and belong to the same universality class specified by Table~\ref{table}.

Finally, we briefly comment on the ground states for $B<B_c$. Numerical results indicates that the entropy density decreases very slowly at low temperatures and appears to saturate to a finite residual value as $T\to0$ (see the Inset in Fig.~\ref{fig:entropy}\textbf{c}). A direct test of this conclusion would require determining the zero-temperature geometries for $B<B_c$. Even though the semiclassical analysis herein suggests a nonzero residual entropy, it is expected that quantum corrections---particularly those arising from low energy gravitational fluctuations---will ultimately restore a vanishing ground-state entropy, similar to the extremal and near-extremal Reissner–Nordstr\"om black holes~\cite{Iliesiu:2020qvm,Ghosh:2019rcj}.

\section{Quantum phase transition at $k=2/3$ with $z=2$}\label{app:z2}
\begin{figure}[h]
\begin{center}
\includegraphics[width=1.0\textwidth]{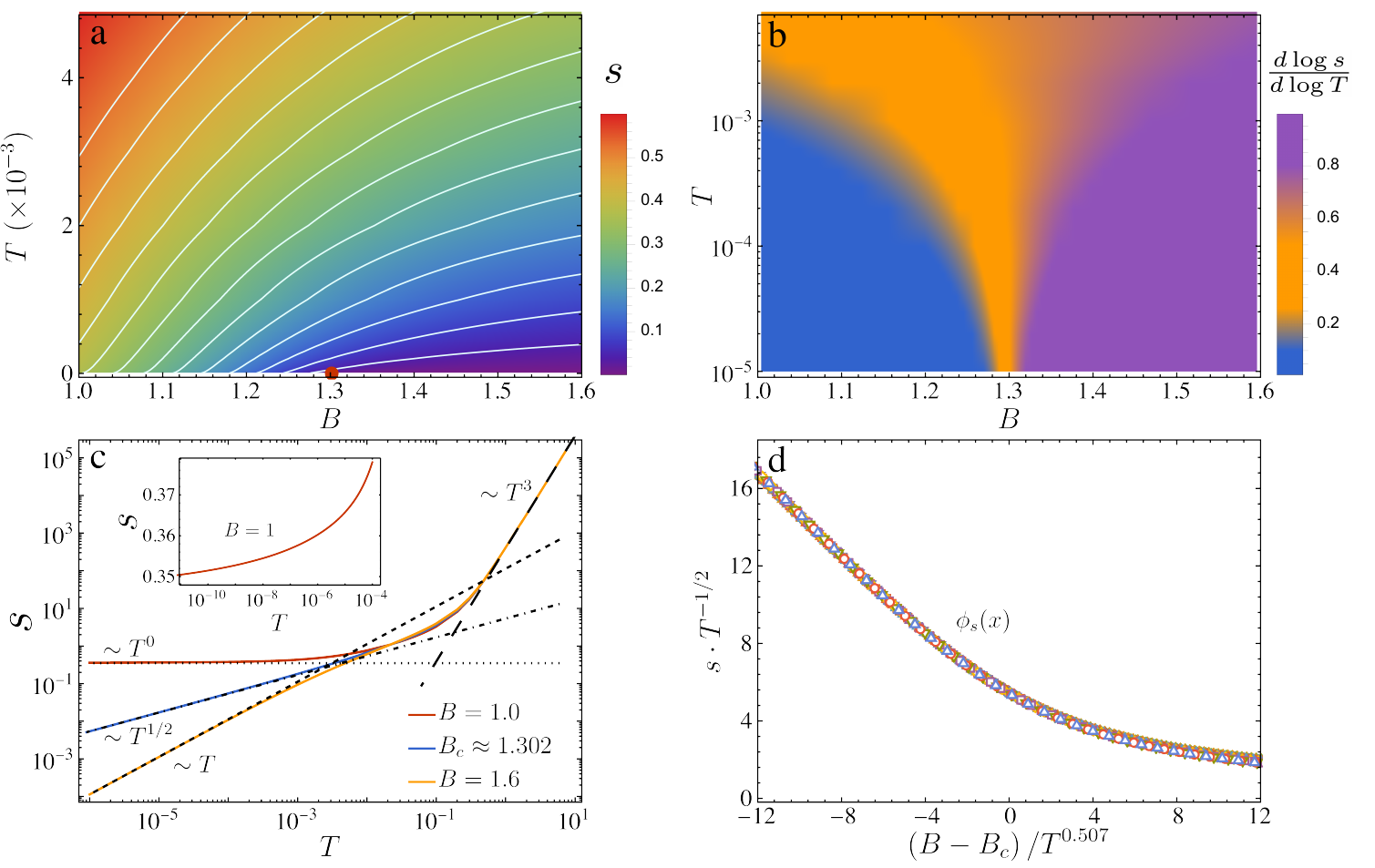}
\end{center}
\vspace{-0.3cm}
\caption{\textbf{Entropy density $s$ and its quantum critical behavior.}
\textbf{a} Density plot of $s$ as functions of magnetic field and temperature for $k=2/3$ and $z=2$. The LightCyan contours represent the isentropes and increase from right to left in steps of $0.1$. The magenta dot marks the position of QCP. 
\textbf{b} Reconstruction of the QCR by computing the regimes where the entropy density $s$ shows a power law scaling.
\textbf{c} Temperature dependence of entropy density in different phases. The inset illustrates the entropy density at extremely low temperatures in the fractionalized phase.
\textbf{d} Data collapse of the entropy density near the QCP within the temperature range $8\times10^{-6}\leq T\leq 10^{-4}$. The plots are for $k=2/3$ and $\mu=1$.}\label{fig:s2}
\end{figure}

Here, we present results for the $z=2$ QCP, which corresponds to the Chern-Simons coupling $k=2/3$ in the action~\eqref{eq:action}. Fig.~\ref{fig:s2}\textbf{a} displays contour plots of the entropy density $s$ in the $T-B$ plane. The QCR appears as the orange region in Fig.~\ref{fig:s2}\textbf{b}. The QCP itself, indicated by the red dot in Fig.~\ref{fig:s2}\textbf{a}, is located at $B=B_c\approx 1.30165$. At the QCP, the entropy density follows the scaling form $s\propto T^{1/2}$, as shown in Fig.~\ref{fig:s2}\textbf{c}. For comparison, Fig.~\ref{fig:s2}\textbf{c} also includes the temperature dependence of the entropy density in the two phases---the cohesive and fractionalized phases---corresponding to the blue and purple regions in Fig.~\ref{fig:s2}\textbf{b}. Finally, the universal scaling function of the entropy density is obtained via data collapse, as demonstrated in Fig.~\ref{fig:s2}\textbf{d}. Following standard scaling analysis, the dynamical exponent is determined to be $z=2d$. Since the near-horizon geometry at the QCP is governed by a three-dimensional ``Schrödinger" spacetime, this suggests that, via holography, the effective spatial dimension of the dual system is $d=1$. As a result, $z=2$ for the Chern-Simons coupling $k=2/3$. Furthermore, data collapse of the entropy density yields $1/z\nu\approx0.507$, from which the correlation length exponent is estimated to be $\nu\approx0.986$.

\begin{figure}[h]
\begin{center}
\includegraphics[width=0.85\textwidth]{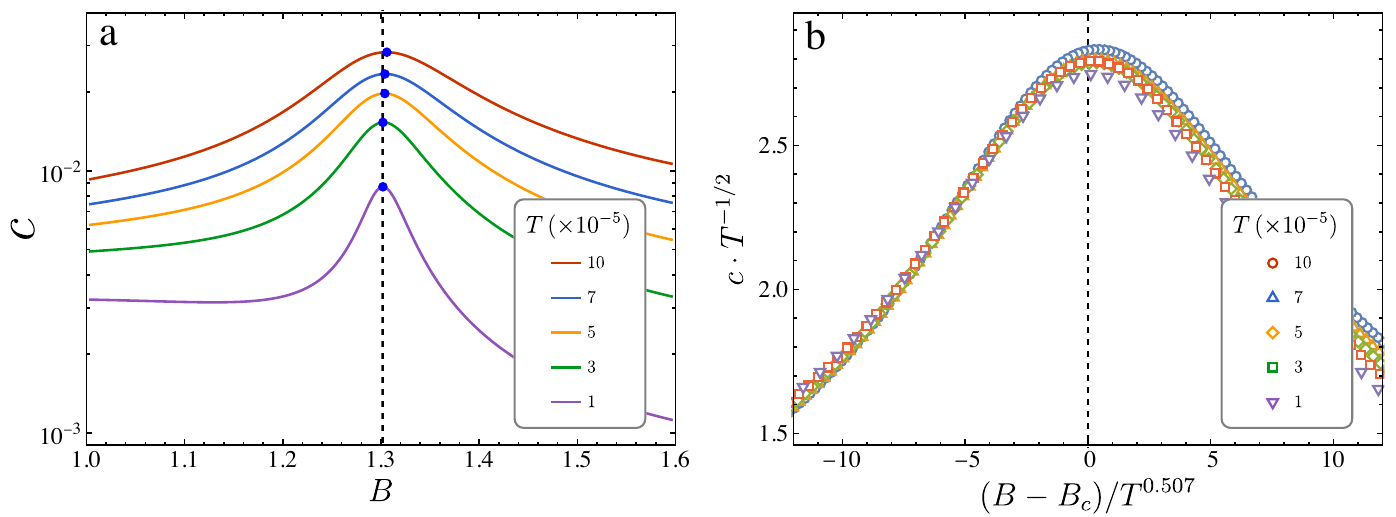}
\end{center}
\vspace{-0.3cm}
\caption{\textbf{Specific heat}. \textbf{a} Magnetic field dependence of specific heat $C$ at different temperatures. The vertical dashed black line marks the location of QCP. The blue point marks the maximum of the specific heat. \textbf{b} Data collapse of the specific heat near the QCR. The plots are for $k=2/3$ and $\mu=1$.}
\label{fig:sheat2}
\end{figure}

\begin{figure}[h]
\begin{center}
\includegraphics[width=1.0\textwidth]{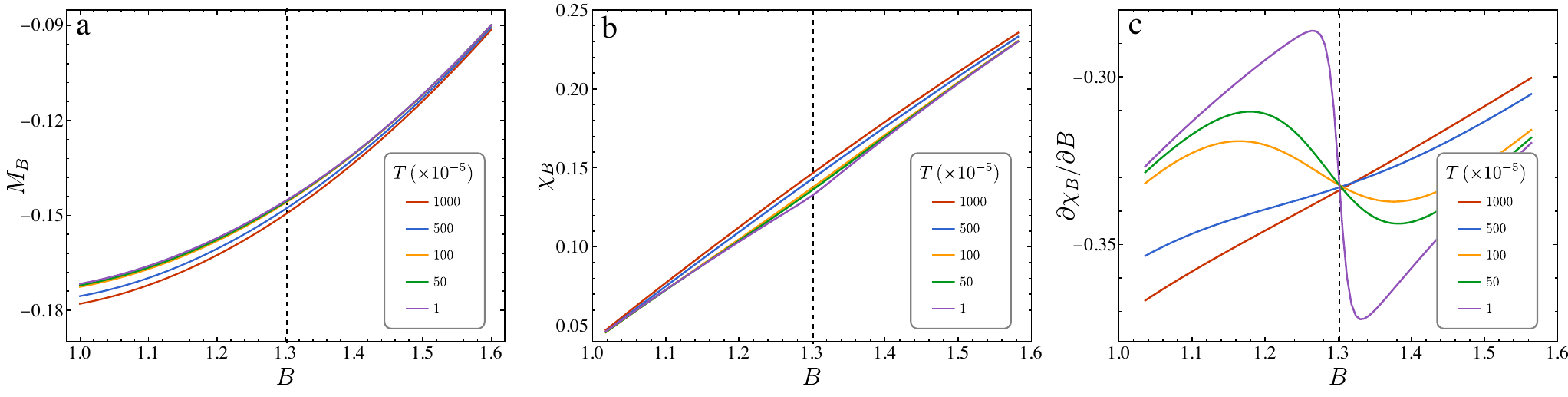}
\end{center}
\vspace{-0.3cm}
\caption{\textbf{Magnetic properties of the QPT.}~\textbf{a} Magnetic field dependence of magnetization $M_B$ at different temperatures. The vertical dashed line marks the location of QCP.~\textbf{b} Behavior of magnetic susceptibility $\chi_B$ as a function of magnetic field $B$ for various temperatures.~\textbf{c} The behavior of $\partial \chi_B/\partial_B$ with respect to $B$. The plots are for $k=2/3$ and $\mu=1$.}
\label{fig:z2-mBchi}
\end{figure}

The behavior of the specific heat density $c$ as a function of magnetic field at various temperatures is shown in Fig.~\ref{fig:sheat2}\textbf{a}. The peaks of the specific heat curves, indicated by blue dots, shift progressively toward the critical point with the decreasing of temperature. Fig.~\ref{fig:sheat2}\textbf{b} demonstrates data collapse of the specific heat within the QCR, where all data fall neatly onto a universal scaling curve. Similar to the $z=3$ case (see Fig.~\ref{fig:sheat}\textbf{b}), there is only one peak in the universal scaling function of specific heat. Fig.~\ref{fig:z2-mBchi}\textbf{a-b} shows the behavior of magnetization and magnetic susceptibility as a function of the magnetic field at different temperatures. Both the magnetization and magnetic susceptibility vary continuously with the magnetic field across the QCP. Since the magnetic susceptibility $\chi_B$ remains positive for magnetic field above and below $B_c$, this indicates that both phases are paramagnetic. Furthermore, as shown in Fig.~\ref{fig:z2-mBchi}\textbf{c}, we find that the derivative of the magnetic susceptibility with respect to $B$ becomes discontinuous at low temperatures, signaling a third order QPT. Given that this transition is of third order for $k=2/3$, the standard scaling relations no longer apply. As a result, only $z$ and $\nu$ can be extracted, and the full set of critical exponents cannot be determined.

\begin{figure}[h]
\begin{center}
\includegraphics[width=0.85\textwidth]{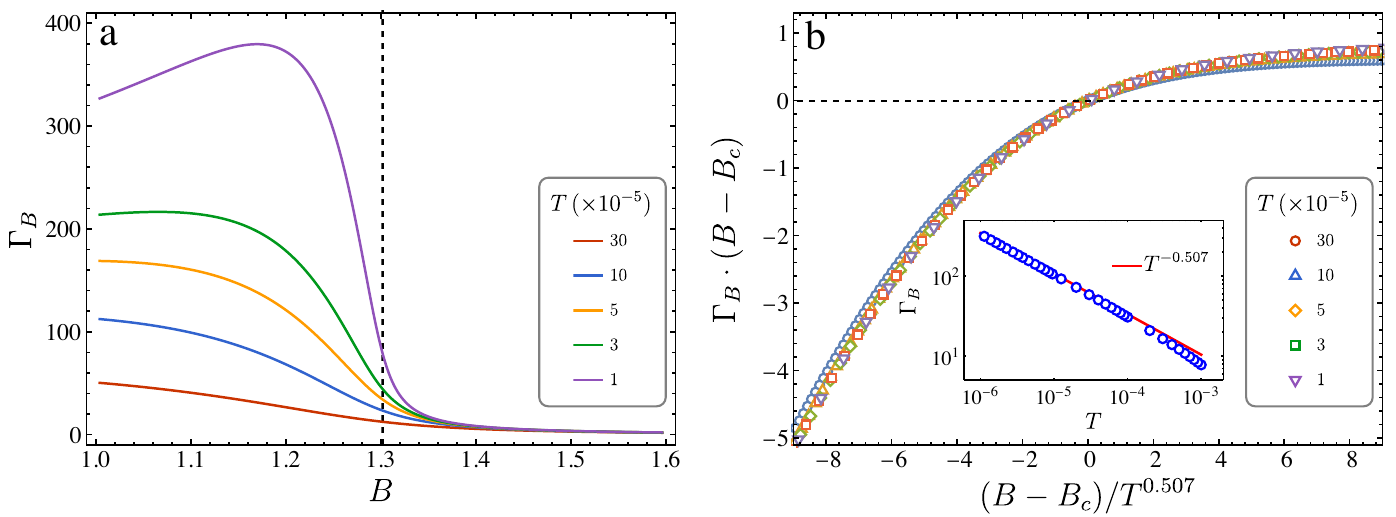}
\end{center}
\vspace{-0.3cm}
\caption{\textbf{Gr\"uneisen ratio of the holographic quantum critically}.~\textbf{a} Magnetic field dependence of the magnetic Gr\"uneisen ratio $\Gamma_B$ across the QPT.~\textbf{b} Data collapse of Gr\"uneisen ratio. The insert fig represents the temperature scaling of $\Gamma_B$ at $B=B_c$. The plots are for $k=2/3$ and $\mu=1$.} \label{fig:z2-Gr}
\end{figure}

Having investigated the thermodynamic properties of the system, we now turn to the novel MCE associated with this quantum criticality. Fig.~\ref{fig:z2-Gr}\textbf{a} shows the behavior of the Gr\"uneisen ratio $\Gamma_B$ as a function of the magnetic field at various temperatures. As the temperature decreases, $\Gamma_B$ is observed to diverge near QCP. Notably, $\Gamma_B$ remains positive throughout the transition, indicating the absence of the peak-dip structure and sign change typically observed in most QPTs. Finally, Fig.~\ref{fig:z2-Gr}\textbf{b} demonstrates the data collapse of the Gr\"uneisen ratio. The $\Gamma_B(B-B_c)$, when plotted as a function of $x=(B-B_c)/T^{0.507}$, collapse onto a same universal curve, confirming the universal scaling behavior of the Gr\"uneisen ratio. Moreover, along $B=B_c$ (or $x=0$), the Gr\"uneisen ratio exhibits the expected scaling behavior $\Gamma_B  \propto T^{-1/z\nu}= T^{-0.507}$. Therefore, we find that the Gr\"uneisen ratio continues to exhibit divergent behavior even for a third-order QPT, and its temperature dependence at the QCP provides an effective means to extract the product $z\nu$ and hence $\nu$. Remarkably, the value of $z\nu$ obtained from $\Gamma_B$ is consistent with that deduced from the data collapse of the entropy and specific heat. This agreement suggests that the universal scaling laws governing the divergence of the Gr\"uneisen ratio extends to third order QPTs as well.

\bibliographystyle{unsrt}
\bibliography{hMQC}

\begin{thebibliography}{10}

\bibitem{Sondhi1997}
S.~L. Sondhi, S.~M. Girvin, J.~P. Carini, and D.~Shahar.
\newblock Continuous quantum phase transitions.
\newblock {\em Rev. Mod. Phys.}, 69:315--333, Jan 1997.

\bibitem{sachdev2015}
Subir Sachdev.
\newblock {\em Quantum phase transitions}.
\newblock Cambridge University Press, Cambridge, second edition, 2015.

\bibitem{sachdev2000}
Subir Sachdev.
\newblock Quantum criticality: Competing ground states in low dimensions.
\newblock {\em Science}, 288(5465):475--480, 2000.

\bibitem{Coleman2005}
Piers Coleman and Andrew~J. Schofield.
\newblock Quantum criticality.
\newblock {\em Nature}, 433:226--229, 2005.

\bibitem{Sebastian2006Purple}
S.~E. Sebastian, N.~Harrison, C.~D. Batista, L.~Balicas, M.~Jaime, P.~A. Sharma, N.~Kawashima, and I.~R. Fisher.
\newblock Dimensional reduction at a quantum critical point.
\newblock {\em Nature}, 441(7093):617--620, 2006.

\bibitem{Gegenwart2007}
P.~Gegenwart, T.~Westerkamp, C.~Krellner, Y.~Tokiwa, S.~Paschen, C.~Geibel, F.~Steglich, E.~Abrahams, and Q.~Si.
\newblock Multiple energy scales at a quantum critical point.
\newblock {\em Science}, 315(5814):969--971, 2007.

\bibitem{coldea2010}
R.~Coldea, D.~A. Tennant, E.~M. Wheeler, E.~Wawrzynska, D.~Prabhakaran, M.~Telling, K.~Habicht, P.~Smeibidl, and K.~Kiefer.
\newblock Quantum criticality in an {Ising} chain: Experimental evidence for emergent ${E}_8$ symmetry.
\newblock {\em Science}, 327(5962):177--180, 2010.

\bibitem{Cubrovic:2009ye}
Mihailo Cubrovic, Jan Zaanen, and Koenraad Schalm.
\newblock {String Theory, Quantum Phase Transitions and the Emergent Fermi-Liquid}.
\newblock {\em Science}, 325:439--444, 2009.

\bibitem{Xiang2024GiantME}
Junsen Xiang, Chuandi Zhang, Yuan Gao, Wolfgang Schmidt, Karin Schmalzl, Chin-Wei Wang, Bo~Li, Ning Xi, Xin-Yang Liu, Hai Jin, Gang Li, Jun Shen, Ziyu Chen, Yang Qi, Yuan Wan, Wentao Jin, Wei Li, Peijie Sun, and Gang Su.
\newblock Giant magnetocaloric effect in spin supersolid candidate {Na$_2$BaCo(PO$_4$)$_2$}.
\newblock {\em Nature}, 625 7994:270--275, 2024.

\bibitem{Gruneisen1912}
E.~Grüneisen.
\newblock Theorie des festen zustandes einatomiger elemente.
\newblock {\em Ann. Phys.}, 344(12):257--306, 1912.

\bibitem{Wolf2014CoolingTQ}
Bernd Wolf, Andreas Honecker, Walter Hofstetter, Ulrich Tutsch, and Michael Lang.
\newblock Cooling through quantum criticality and many-body effects in condensed matter and cold gases.
\newblock {\em arXiv: Strongly Correlated Electrons}, 2014.

\bibitem{Shu2026Nature}
Mingfang Shu, Xitong Xu, Ning Xi, Miao He, Junsen Xiang, Gexing Qu, Dmitry Khalyavin, Pascal Manuel, Jumpei~G. Nakamura, Jinlong Jiao, Yonglai Liu, Guoliang Wu, Kaizhen Guo, Haitian Zhao, Wei Xu, Qingchen Duan, Ruidan Zhong, Xinqing Wang, Yuyan Han, Langsheng Ling, Xuefeng Sun, Dongsheng Song, Yuan Gao, Zhentao Wang, Xi~Chen, Tian Qian, Shuang Jia, Haifeng Du, Gang Su, Wei Li, Jie Ma, and Zhe Qu.
\newblock Giant magnetocaloric effect and spin supersolid in a metallic dipolar magnet.
\newblock {\em Nature}, 651(8104):61--67, 2026.

\bibitem{Zhu2003UniversallyDG}
Lijun Zhu, Markus Garst, Achim Rosch, and Qimiao Si.
\newblock Universally diverging gr{\"u}neisen parameter and the magnetocaloric effect close to quantum critical points.
\newblock {\em Physical review letters}, 91 6:066404, 2003.

\bibitem{Garst2005SignCO}
Markus Garst and Achim Rosch.
\newblock Sign change of the gruneisen parameter and magnetocaloric effect near quantum critical points.
\newblock {\em Physical Review B}, 72:205129, 2005.

\bibitem{sciadvaao3773}
Oliver Breunig, Markus Garst, Andreas Klümper, Jens Rohrkamp, Mark~M. Turnbull, and Thomas Lorenz.
\newblock Quantum criticality in the spin-1/2 heisenberg chain system copper pyrazine dinitrate.
\newblock {\em Science Advances}, 3(12):eaao3773, 2017.

\bibitem{xiang2025}
Junsen Xiang, Enze Lv, Qinxin Shen, Cheng Su, Xuetong He, Yinghao Zhu, Yuan Gao, Xin-Yang Liu, Dai-Wei Qu, Xinlei Wang, Xi~Chen, Qian Zhao, Haifeng Li, Shuo Li, Jie Yang, Jun Luo, Peijie Sun, Wentao Jin, Yang Qi, Rui Zhou, Wei Li, and Gang Su.
\newblock Universal magnetocaloric effect near quantum critical point of magnon {B}ose-{E}instein condensation, 2025.

\bibitem{Kuchler2003}
R.~K\"uchler, N.~Oeschler, P.~Gegenwart, T.~Cichorek, K.~Neumaier, O.~Tegus, C.~Geibel, J.~A. Mydosh, F.~Steglich, L.~Zhu, and Q.~Si.
\newblock Divergence of the {G}r\"uneisen ratio at quantum critical points in heavy fermion metals.
\newblock {\em Phys. Rev. Lett.}, 91:066405, Aug 2003.

\bibitem{Kuchler2004}
R.~K\"uchler, P.~Gegenwart, K.~Heuser, E.-W. Scheidt, G.~R. Stewart, and F.~Steglich.
\newblock Gr\"uneisen ratio divergence at the quantum critical point in $\mathrm{CeCu}_{6-x}\ \mathrm{Ag}_{x}$.
\newblock {\em Phys. Rev. Lett.}, 93:096402, Aug 2004.

\bibitem{Kuchler_2007}
R.~Küchler, P.~Gegenwart, C.~Geibel, and F.~Steglich.
\newblock Systematic study of the {G}rüneisen ratio near quantum critical points.
\newblock {\em Science and Technology of Advanced Materials}, 8(5):428, jul 2007.

\bibitem{Tokiwa2009}
Y.~Tokiwa, T.~Radu, C.~Geibel, F.~Steglich, and P.~Gegenwart.
\newblock {Divergence of the magnetic Gr\"uneisen ratio at the field-induced quantum critical point in {YbRh$_{2}$Si$_{2}$}}.
\newblock {\em Phys. Rev. Lett.}, 102:066401, Feb 2009.

\bibitem{Alexander2013}
Alexander Steppke, Robert Küchler, Stefan Lausberg, Edit Lengyel, Lucia Steinke, Robert Borth, Thomas Lühmann, Cornelius Krellner, Michael Nicklas, Christoph Geibel, Frank Steglich, and Manuel Brando.
\newblock Ferromagnetic quantum critical point in the heavy-fermion metal {YbNi$_4$(P$_{1-x}$As$_x$)$_2$}.
\newblock {\em Science}, 339(6122):933--936, 2013.

\bibitem{Gegenwart2016GrneisenPS}
Philipp Gegenwart.
\newblock Gr{\"u}neisen parameter studies on heavy fermion quantum criticality.
\newblock {\em Reports on Progress in Physics}, 79, 2016.

\bibitem{Bin2020}
Bin Shen, Yongjun Zhang, Yashar Komijani, Michael Nicklas, Robert Borth, An~Wang, Ye~Chen, Zhiyong Nie, Rui Li, Xin Lu, Hanoh Lee, Michael Smidman, Frank Steglich, Piers Coleman, and Huiqiu Yuan.
\newblock Strange-metal behaviour in a pure ferromagnetic {K}ondo lattice.
\newblock {\em Nature}, 579:51--55, 2020.

\bibitem{Zhan2025CriticalFA}
Jin Zhan, Yongjun Zhang, Jiawen Zhang, Yu~Liu, Zhiyong Nie, Yuxin Chen, Lin Jiao, Yashar Komijani, Michael Smidman, Frank Steglich, Piers Coleman, and Huiqiu Yuan.
\newblock Critical fluctuations and conserved dynamics in a strange ferromagnetic metal.
\newblock {\em Physical review letters}, 135 26:266504, 2025.

\bibitem{Hertz:1976zz}
John~A. Hertz.
\newblock Quantum critical phenomena.
\newblock {\em Phys. Rev. B}, 14:1165--1184, Aug 1976.

\bibitem{Millis:1993}
A.~J. Millis.
\newblock Effect of a nonzero temperature on quantum critical points in itinerant fermion systems.
\newblock {\em Phys. Rev. B}, 48:7183--7196, Sep 1993.

\bibitem{Lohneysen:2007zz}
Hilbert~v. Lohneysen, Achim Rosch, Matthias Vojta, and Peter Wolfle.
\newblock {Fermi-liquid instabilities at magnetic quantum phase transitions}.
\newblock {\em Rev. Mod. Phys.}, 79:1015--1075, 2007.

\bibitem{Maldacena:1997re}
Juan~Martin Maldacena.
\newblock {The Large $N$ limit of superconformal field theories and supergravity}.
\newblock {\em Adv. Theor. Math. Phys.}, 2:231--252, 1998.

\bibitem{Witten:1998qj}
Edward Witten.
\newblock {Anti de Sitter space and holography}.
\newblock {\em Adv. Theor. Math. Phys.}, 2:253--291, 1998.

\bibitem{Gubser:1998bc}
S.~S. Gubser, Igor~R. Klebanov, and Alexander~M. Polyakov.
\newblock {Gauge theory correlators from noncritical string theory}.
\newblock {\em Phys. Lett. B}, 428:105--114, 1998.

\bibitem{Liu:2009dm}
Hong Liu, John McGreevy, and David Vegh.
\newblock {Non-Fermi liquids from holography}.
\newblock {\em Phys. Rev. D}, 83:065029, 2011.

\bibitem{Faulkner:2009wj}
Thomas Faulkner, Hong Liu, John McGreevy, and David Vegh.
\newblock {Emergent quantum criticality, Fermi surfaces, and AdS(2)}.
\newblock {\em Phys. Rev. D}, 83:125002, 2011.

\bibitem{Horowitz:2012ky}
Gary~T. Horowitz, Jorge~E. Santos, and David Tong.
\newblock {Optical Conductivity with Holographic Lattices}.
\newblock {\em JHEP}, 07:168, 2012.

\bibitem{Sachdev1992GaplessSG}
Subir Sachdev and Jinwu Ye.
\newblock Gapless spin-fluid ground state in a random quantum heisenberg magnet.
\newblock {\em Physical review letters}, 70 21:3339--3342, 1992.

\bibitem{Maldacena:2016hyu}
Juan Maldacena and Douglas Stanford.
\newblock {Remarks on the Sachdev-Ye-Kitaev model}.
\newblock {\em Phys. Rev. D}, 94(10):106002, 2016.

\bibitem{Maldacena:2016upp}
Juan Maldacena, Douglas Stanford, and Zhenbin Yang.
\newblock {Conformal symmetry and its breaking in two dimensional Nearly Anti-de-Sitter space}.
\newblock {\em PTEP}, 2016(12):12C104, 2016.

\bibitem{Hartnoll:2008vx}
Sean~A. Hartnoll, Christopher~P. Herzog, and Gary~T. Horowitz.
\newblock {Building a Holographic Superconductor}.
\newblock {\em Phys. Rev. Lett.}, 101:031601, 2008.

\bibitem{Adams:2012pj}
Allan Adams, Paul~M. Chesler, and Hong Liu.
\newblock {Holographic Vortex Liquids and Superfluid Turbulence}.
\newblock {\em Science}, 341:368--372, 2013.

\bibitem{Cai:2017qdz}
Rong-Gen Cai, Li~Li, Yong-Qiang Wang, and Jan Zaanen.
\newblock {Intertwined Order and Holography: The Case of Parity Breaking Pair Density Waves}.
\newblock {\em Phys. Rev. Lett.}, 119(18):181601, 2017.

\bibitem{Chesler:2008hg}
Paul~M. Chesler and Laurence~G. Yaffe.
\newblock {Horizon formation and far-from-equilibrium isotropization in supersymmetric Yang-Mills plasma}.
\newblock {\em Phys. Rev. Lett.}, 102:211601, 2009.

\bibitem{Adams:2013vsa}
Allan Adams, Paul~M. Chesler, and Hong Liu.
\newblock {Holographic turbulence}.
\newblock {\em Phys. Rev. Lett.}, 112(15):151602, 2014.

\bibitem{bhaseen2015}
M.~J. Bhaseen, Benjamin Doyon, Andrew Lucas, and Koenraad Schalm.
\newblock Energy flow in quantum critical systems far from equilibrium.
\newblock {\em Nature Physics}, 11(6):509--514, June 2015.

\bibitem{Baggioli:2021tzr}
Matteo Baggioli, Li~Li, and Hao-Tian Sun.
\newblock {Shear Flows in Far-from-Equilibrium Strongly Coupled Fluids}.
\newblock {\em Phys. Rev. Lett.}, 129(1):011602, 2022.

\bibitem{Ryu:2006bv}
Shinsei Ryu and Tadashi Takayanagi.
\newblock {Holographic derivation of entanglement entropy from AdS/CFT}.
\newblock {\em Phys. Rev. Lett.}, 96:181602, 2006.

\bibitem{Takayanagi:2017knl}
Tadashi Takayanagi and Koji Umemoto.
\newblock {Entanglement of purification through holographic duality}.
\newblock {\em Nature Phys.}, 14(6):573--577, 2018.

\bibitem{Takayanagi:2025ula}
Tadashi Takayanagi.
\newblock {Essay: Emergent Holographic Spacetime from Quantum Information}.
\newblock {\em Phys. Rev. Lett.}, 134(24):240001, 2025.

\bibitem{Zaanen:2015oix}
Jan Zaanen, Ya-Wen Sun, Yan Liu, and Koenraad Schalm.
\newblock {\em {Holographic Duality in Condensed Matter Physics}}.
\newblock Cambridge Univ. Press, 2015.

\bibitem{Ammon:2015wua}
Martin Ammon and Johanna Erdmenger.
\newblock {\em {Gauge/gravity duality}: {Foundations and applications}}.
\newblock Cambridge University Press, Cambridge, 4 2015.

\bibitem{Cai:2015cya}
Rong-Gen Cai, Li~Li, Li-Fang Li, and Run-Qiu Yang.
\newblock {Introduction to Holographic Superconductor Models}.
\newblock {\em Sci. China Phys. Mech. Astron.}, 58(6):060401, 2015.

\bibitem{Hartnoll:2018xxg}
Sean~A. Hartnoll, Andrew Lucas, and Subir Sachdev.
\newblock {\em {Holographic Quantum Matter}}.
\newblock MIT Press, 2018.

\bibitem{Baggioli:2021xuv}
Matteo Baggioli, Keun-Young Kim, Li~Li, and Wei-Jia Li.
\newblock {Holographic Axion Model: a simple gravitational tool for quantum matter}.
\newblock {\em Sci. China Phys. Mech. Astron.}, 64(7):270001, 2021.

\bibitem{Hartnoll:2011pp}
Sean~A. Hartnoll and Liza Huijse.
\newblock {Fractionalization of holographic Fermi surfaces}.
\newblock {\em Class. Quant. Grav.}, 29:194001, 2012.

\bibitem{Hartnoll:2012ux}
Sean~A. Hartnoll and Djordje Radicevic.
\newblock {Holographic order parameter for charge fractionalization}.
\newblock {\em Phys. Rev. D}, 86:066001, 2012.

\bibitem{Donos:2012js}
Aristomenis Donos and Sean~A. Hartnoll.
\newblock {Interaction-driven localization in holography}.
\newblock {\em Nature Phys.}, 9:649--655, 2013.

\bibitem{Landsteiner:2015pdh}
Karl Landsteiner, Yan Liu, and Ya-Wen Sun.
\newblock {Quantum phase transition between a topological and a trivial semimetal from holography}.
\newblock {\em Phys. Rev. Lett.}, 116(8):081602, 2016.

\bibitem{Ammon:2016mwa}
Martin Ammon, Markus Heinrich, Amadeo Jim{\'e}nez-Alba, and Sebastian Moeckel.
\newblock {Surface States in Holographic Weyl Semimetals}.
\newblock {\em Phys. Rev. Lett.}, 118(20):201601, 2017.

\bibitem{Liu:2018spp}
Yan Liu and Junkun Zhao.
\newblock {Weyl semimetal/insulator transition from holography}.
\newblock {\em JHEP}, 12:124, 2018.

\bibitem{Erdmenger:2008rm}
Johanna Erdmenger, Michael Haack, Matthias Kaminski, and Amos Yarom.
\newblock {Fluid dynamics of R-charged black holes}.
\newblock {\em JHEP}, 01:055, 2009.

\bibitem{Banerjee:2008th}
Nabamita Banerjee, Jyotirmoy Bhattacharya, Sayantani Bhattacharyya, Suvankar Dutta, R.~Loganayagam, and P.~Surowka.
\newblock {Hydrodynamics from charged black branes}.
\newblock {\em JHEP}, 01:094, 2011.

\bibitem{Son:2009tf}
Dam~T. Son and Piotr Surowka.
\newblock {Hydrodynamics with Triangle Anomalies}.
\newblock {\em Phys. Rev. Lett.}, 103:191601, 2009.

\bibitem{Donos:2012wi}
Aristomenis Donos and Jerome~P. Gauntlett.
\newblock {Black holes dual to helical current phases}.
\newblock {\em Phys. Rev. D}, 86:064010, 2012.

\bibitem{Rangamani:2023mok}
Mukund Rangamani, Julio Virrueta, and Shuyan Zhou.
\newblock {Anomalous hydrodynamics effective actions from holography}.
\newblock {\em JHEP}, 11:044, 2023.

\bibitem{DHoker:2010zpp}
Eric D'Hoker and Per Kraus.
\newblock {Holographic Metamagnetism, Quantum Criticality, and Crossover Behavior}.
\newblock {\em JHEP}, 05:083, 2010.

\bibitem{DHoker:2010onp}
Eric D'Hoker and Per Kraus.
\newblock {Magnetic Field Induced Quantum Criticality via new Asymptotically AdS$_{5}$ Solutions}.
\newblock {\em Class. Quant. Grav.}, 27:215022, 2010.

\bibitem{DHoker:2012rlj}
Eric D'Hoker and Per Kraus.
\newblock {Quantum Criticality via Magnetic Branes}.
\newblock {\em Lect. Notes Phys.}, 871:469--502, 2013.

\bibitem{Buchel:2006gb}
Alex Buchel and James~T. Liu.
\newblock {Gauged supergravity from type IIB string theory on Y**p,q manifolds}.
\newblock {\em Nucl. Phys. B}, 771:93--112, 2007.

\bibitem{Gauntlett:2006ai}
Jerome~P. Gauntlett, Eoin O~Colgain, and Oscar Varela.
\newblock {Properties of some conformal field theories with M-theory duals}.
\newblock {\em JHEP}, 02:049, 2007.

\bibitem{Zhao:2025gej}
Jun-Kun Zhao and Li~Li.
\newblock {Holographic study of shear viscosity and butterfly velocity for magnetic field-driven quantum criticality}.
\newblock {\em JHEP}, 10:131, 2025.

\bibitem{Son:2008ye}
D.~T. Son.
\newblock {Toward an AdS/cold atoms correspondence: A Geometric realization of the Schrodinger symmetry}.
\newblock {\em Phys. Rev. D}, 78:046003, 2008.

\bibitem{Balasubramanian:2008dm}
Koushik Balasubramanian and John McGreevy.
\newblock {Gravity duals for non-relativistic CFTs}.
\newblock {\em Phys. Rev. Lett.}, 101:061601, 2008.

\bibitem{Channarayappa2024}
Sharath~Kumar Channarayappa, Sankalp Kumar, N~S Vidhyadhiraja, Sumiran Pujari, M~P Saravanan, Amal Sebastian, Eun~Sang Choi, Shalinee Chikara, Dolly Nambi, Athira Suresh, Siddhartha Lal, and D~Jaiswal-Nagar.
\newblock {Tomonaga–Luttinger} liquid and quantum criticality in spin-1/2 antiferromagnetic {H}eisenberg chain {C}$_{14}${H}$_{18}${C}u{N}$_4${O}$_{10}$ via {W}ilson ratio.
\newblock {\em PNAS Nexus}, 3(9):pgae363, 08 2024.

\bibitem{Millis2001MetamagneticQC}
Andrew~J. Millis, Andrew~J. Schofield, Gilbert~George Lonzarich, and Santiago~A. Grigera.
\newblock Metamagnetic quantum criticality in metals.
\newblock {\em Physical review letters}, 88 21:217204, 2001.

\bibitem{Continentino_2017}
Mucio Continentino.
\newblock {\em Quantum Scaling in Many-Body Systems: An Approach to Quantum Phase Transitions}.
\newblock Cambridge University Press, 2 edition, 2017.

\bibitem{2025arXiv250917362Z}
Xuan {Zhou}, Enze {Lv}, Wei {Li}, and Yang {Qi}.
\newblock {Universal Scaling Functions of the Gr\{{\"u}\}neisen Ratio near Quantum Critical Points}.
\newblock {\em arXiv e-prints}, page arXiv:2509.17362, September 2025.

\bibitem{Senthil:2002vtt}
T.~Senthil, Subir Sachdev, and Matthias Vojta.
\newblock {Fractionalized Fermi liquids}.
\newblock {\em Phys. Rev. Lett.}, 90(21):216403, 2003.

\bibitem{Ammon:2016szz}
Martin Ammon, Julian Leiber, and Rodrigo~P. Macedo.
\newblock {Phase diagram of 4D field theories with chiral anomaly from holography}.
\newblock {\em JHEP}, 03:164, 2016.

\bibitem{Nakamura:2009tf}
Shin Nakamura, Hirosi Ooguri, and Chang-Soon Park.
\newblock {Gravity Dual of Spatially Modulated Phase}.
\newblock {\em Phys. Rev. D}, 81:044018, 2010.

\bibitem{Basar:2010zd}
Gokce Basar, Gerald~V. Dunne, and Dmitri~E. Kharzeev.
\newblock {Chiral Magnetic Spiral}.
\newblock {\em Phys. Rev. Lett.}, 104:232301, 2010.

\bibitem{Fradkin:2010ARCMP}
Eduardo {Fradkin}, Steven~A. {Kivelson}, Michael~J. {Lawler}, James~P. {Eisenstein}, and Andrew~P. {MacKenzie}.
\newblock {Nematic Fermi Fluids in Condensed Matter Physics}.
\newblock {\em Annual Review of Condensed Matter Physics}, 1:153--178, April 2010.

\bibitem{Metlitski:2010pd}
Max~A. Metlitski and Subir Sachdev.
\newblock {Quantum phase transitions of metals in two spatial dimensions: I. Ising-nematic order}.
\newblock {\em Phys. Rev. B}, 82:075127, 2010.

\bibitem{boyd2001chebyshev}
John~P Boyd.
\newblock {\em Chebyshev and Fourier spectral methods}.
\newblock Courier Corporation, 2001.

\bibitem{Cai:2024tyv}
Rong-Gen Cai, Li~Li, and Jun-Kun Zhao.
\newblock {Thermodynamics of dyonic black holes in minimal supergravity}.
\newblock 10 2024.

\bibitem{Iliesiu:2020qvm}
Luca~V. Iliesiu and Gustavo~J. Turiaci.
\newblock {The statistical mechanics of near-extremal black holes}.
\newblock {\em JHEP}, 05:145, 2021.

\bibitem{Ghosh:2019rcj}
Animik Ghosh, Henry Maxfield, and Gustavo~J. Turiaci.
\newblock {A universal Schwarzian sector in two-dimensional conformal field theories}.
\newblock {\em JHEP}, 05:104, 2020.

\end{thebibliography}

\end{document}